\documentclass[
  twocolumn,
  pra,
  amsmath,
  amssymb,
  superscriptaddress
]{revtex4}

\usepackage{bm}
\usepackage{graphicx}
\usepackage{color}

\def\be{\begin{equation}}
\def\ee{\end{equation}}
\def\bea{\begin{eqnarray}}
\def\eea{\end{eqnarray}}

\begin{document}

\title{Many-body delocalization transition and relaxation in a quantum dot}

\author{I.~V.\ Gornyi}
\affiliation{Institut f\"ur Nanotechnologie, Karlsruhe Institute of Technology, 76021 Karlsruhe, Germany}
\affiliation{\mbox{Institut f\"ur Theorie der Kondensierten Materie, Karlsruhe Institute of Technology, 76128 Karlsruhe, Germany}}
\affiliation{A.F.\ Ioffe Physico-Technical Institute, 194021 St.~Petersburg, Russia}
\affiliation{L.D.\ Landau Institute for Theoretical Physics RAS, 119334 Moscow, Russia}

\author{A.~D.\ Mirlin}
\affiliation{Institut f\"ur Nanotechnologie, Karlsruhe Institute of Technology, 76021 Karlsruhe, Germany}
\affiliation{\mbox{Institut f\"ur Theorie der Kondensierten Materie, Karlsruhe Institute of Technology, 76128 Karlsruhe, Germany}}
\affiliation{Petersburg Nuclear Physics Institute, 188300 St.~Petersburg, Russia}
\affiliation{L.D.\ Landau Institute for Theoretical Physics RAS, 119334 Moscow, Russia}

\author{D.~G.\ Polyakov}
\affiliation{Institut f\"ur Nanotechnologie, Karlsruhe Institute of Technology, 76021 Karlsruhe, Germany}

\begin{abstract}
We revisit the problem of quantum localization of many-body states in a quantum dot and the associated problem of relaxation of an excited state in a finite correlated electron system. We determine the localization threshold for the eigenstates in Fock space. We argue that the localization-delocalization transition (which manifests itself, e.g., in the statistics of many-body energy levels) becomes sharp in the limit of a large dimensionless conductance (or, equivalently, in the limit of weak interaction). We also analyze the temporal relaxation of quantum states of various types (a ``hot-electron state'', a ``typical'' many-body state, and a single-electron excitation added to a ``thermal state'') with energies below, at, and above the transition.
\end{abstract}

\maketitle

\section{Introduction}
\label{s0}

Anderson localization~\cite{anderson58} is one of the most fundamental and ubiquitous quantum phenomena. Conventionally, strong localization  is thought of as occurring in real space in spatially extended disordered systems with the system size much larger than the localization length. It was, however, pointed out in a seminal paper~\cite{altshuler97} by Altshuler, Gefen, Kamenev, and Levitov (AGKL) that Anderson localization can also manifest itself in a disordered (or chaotic) quantum dot where single-particle states extend through the whole system. In this case, localization takes place not in the coordinate space but in the Fock space of the interacting quantum system and is closely related to the concepts of ergodicity and thermalization.

The work of AGKL was largely motivated by two influential papers on quantum dots: the experiment by Sivan \textit{et al.}~\cite{sivan94a}, which measured the quasiparticle spectrum of a quantum dot, and by the subsequent theoretical work by Sivan, Imry, and Aronov~\cite{sivan94} (SIA) where a golden-rule analysis of this problem was performed. AGKL emphasized that the golden-rule calculation becomes inapplicable at low quasiparticle energies because of Anderson localization in Fock space. They developed an hierarchical Fock-space model for the problem of a ``hot quasiparticle'' decay in a quantum dot and argued that the problem can be reduced to a tight-binding Anderson model on the Bethe lattice~\cite{abouchacra73,disclaimer1,efetov85,zirnbauer86,efetov87,verbaarschot88,mirlin91} with a large coordination number.
On this basis, they concluded that there is an Anderson localization transition in Fock space which takes place at the energy
of the order of
\be
\label{e1}
E_{1/2}=(g/\ln g)^{1/2}\Delta~.
\ee
Here, $\Delta$ is the characteristic single-electron level spacing in the dot, and $g\gg 1$ is the dimensionless conductance which determines the characteristic value of the interaction matrix elements $V\sim \Delta/g$. The energy scale (\ref{e1}) can be found, up to a logarithmic factor, by equating $V$ and the characteristic level spacing of three-particle states to which a single-particle state is directly coupled by the interaction.

Subsequent work~\cite{jacquod97,mirlin97,silvestrov97,silvestrov98} has corroborated the basic physical picture proposed by AGKL. In particular, the Bethe-lattice framework of Ref.~\cite{altshuler97} was supported and substantiated by the analysis~\cite{mirlin97} of fluctuations near the localization transition on the Bethe lattice with a large coordination number. However, Refs.~\cite{jacquod97,mirlin97,silvestrov97,silvestrov98} also emphasized important deficiencies of the AGKL arguments in regard to the connection between the quantum-dot and Bethe-lattice problems.

First, Jacquod and Shepelyansky~\cite{jacquod97} pointed out
that the number of directly connected states decreases as $1/n$ with increasing generation number $n$ in the problem of a hot quasiparticle decay (in Ref.~\cite{altshuler97}, this circumstance was noted but neglected in order ``to simplify the discussion'') and that this has important ramifications. Namely, Eq.~(\ref{e1}) does not represent the transition energy for the quantum-dot problem but rather gives the lower boundary of a parametrically broad energy interval within which higher Fock-space generations become gradually admixed to the original hot-electron state. One of the questions that arise then is at what energy all generations ``get mixed up''. An extension of the AGKL argument to the resulting ``typical'' many-particle states leads~\cite{jacquod97,mirlin97,silvestrov98} to the replacement of the energy in Eq.~(\ref{e1}) by a parametrically larger energy scale
\be
\label{e2}
E_{2/3}=g^{2/3}\Delta~.
\ee
This energy scale emerges
if one compares $V$ and the level spacing of the states to which a typical state is coupled directly.

Second, Silvestrov~\cite{silvestrov97,silvestrov98} noted strong cancellations in higher orders of the Schr\"odinger perturbation theory
and pointed out~\cite{silvestrov98} that, if the cancellations compensate for the $n!$ growth of the number of contributions to the coupling
at $n$th order, the localization threshold will be at the energy of the order of
\be
\label{e3}
E_1=(g/\ln g)\Delta~,
\ee
instead of Eq.~(\ref{e2}). He stated, however, that this scenario ``seems very unlikely'' and that Eq.~(\ref{e2}) is ``more physically motivated.''

A considerable number of works have studied the problem by means of numerical simulations. In Ref.~\cite{jacquod97}, the scaling analysis of a crossover between the Poisson and Wigner-Dyson statistics of many-body energy levels---and thus between localization and delocalization (quantum chaos)---was performed within the ``two-body random interaction model'' (TBRIM). The conclusion~\cite{jacquod97,[DLS01]} was that the crossover
takes place around $E_{2/3}$.
In Refs.~\cite{mejia-monasterio98} and \cite{leyronas99}, a gradual delocalization of single-particle states within the TBRIM was reported, in qualitative agreement with Refs.~\cite{jacquod97,mirlin97}, and \cite{silvestrov98}. In Refs.~\cite{weinmann97} and \cite{berkovits98}, the localization-delocalization crossover with increasing energy of many-particle states was observed through the exact diagonalization of a tight-binding many-particle Hamiltonian, although the system size was too small to extract information on the scaling of the position and sharpness of the crossover with $g$.

Leyronas, Silvestrov, and Beenakker~\cite{leyronas00} used a restricted (``layered'') version of the TBRIM where only a subset of matrix elements was kept. Analyzing numerically the inverse participation ratio (IPR), which characterizes spreading of many-body eigenstates in Fock space, they concluded that the crossover to chaos takes place around the energy $E_{1}$, and that the crossover likely becomes a sharp transition in the thermodynamic limit. Later, the problem was considered by Rivas, Mucciolo, and Kamenev~\cite{rivas02} within a self-consistent approximation to the quasiparticle Green function. Solving numerically the self-consistent equations, they found a gradual spreading of a hot-electron state starting from the energy scale $E_{1/2}$, in qualitative agreement with the previous analytical arguments~\cite{jacquod97,mirlin97,silvestrov98} and numerical works~\cite{mejia-monasterio98,leyronas99}. On the other hand, the analysis of the IPR for many-body eigenstates led the authors of Ref.~\cite{rivas02} to the conclusion that the localization transition is not governed by the parameter $E/E_1$, contrary to the results of Ref.~\cite{leyronas00}. Specifically, it was argued in Ref.~\cite{rivas02}
that an intermediate regime, where neither the localized-regime nor the golden-rule formulas for the IPR describe the numerical results, becomes broader as $g$ is increased.

More recently, two papers~\cite{gornyi05,basko06} extended the ideas of AGKL to explore the localization properties of high-energy states in disordered many-body systems with spatially localized single-particle states. As a prime example, such a question necessarily arises when one considers (quasi-)one-dimensional systems at nonzero temperature, since all single-particle states are localized in this geometry. It was found~\cite{gornyi05,basko06} that the system exhibits a transition between the low-temperature localized phase~\cite{fleishman80} and the high-temperature delocalized phase~\cite{altshuler81} at the temperature
\be
\label{e4}
T_c \sim \frac{\Delta_\xi}{\alpha\ln \alpha^{-1}}~,
\ee
where $\Delta_\xi$ is the characteristic single-particle level spacing in the localization volume, and $\alpha\ll 1$ is the dimensionless strength of (short-range) interaction. While the approximations used in Refs.~\cite{gornyi05,basko06} were somewhat different, the results for $T_c$ are the same, up to a numerical prefactor, see Ref.~\cite{ros15} for a recent detailed discussion.

References~\cite{gornyi05,basko06} have triggered a considerable amount of research on the nature of many-body localization in extended systems. Most intensively the problem has been studied through numerical simulations of disordered interacting one-dimensional systems, see, in particular, Refs.~\cite{oganesyan07,monthus10,bardarson12,serbyn13,gopalakrishnan14,luitz15,nandkishore15,karrasch15,agarwal15,barlev15}. By and large, the numerical results have provided support for the existence of the localization-delocalization transition. On the other hand, the nature of the delocalized phase still remains controversial. Two recent numerical works~\cite{agarwal15,barlev15} concluded that transport in the delocalized phase just above the transition is of a subdiffusive nature (see also Refs.~\cite{gopalakrishnan15}, \cite{reichmann15}, and \cite{lerose15}). If true, this would imply, at least, an intermediate delocalized phase with zero dc conductivity, with Ref.~\cite{barlev15} suggesting that transport is subdiffusive throughout the whole delocalized part of the parameter space of the problem.

Many-body localization was discussed in both normal and superconducting systems; in particular, Ref.~\cite{feigelman10} analyzed many-body localization in the vicinity of a superconductor-insulator transition, also in connection with available experimental data. Recently, experimental evidence of a finite-temperature many-body localization transition was reported for amorphous indium-oxide films~\cite{ovadyahu,ovadia15} and for fermionic atoms in a ``quasi-random'' one-dimensional optical lattice~\cite{schreiber15} (many-body localization of cold atoms was studied experimentally also in an array of coupled one-dimensional optical lattices~\cite{bordia15}).

In this paper, we revisit the problem of localization of many-body states in a quantum dot. While a large number of works addressed this problem both analytically and numerically (mainly during the years 1997-2002, as outlined above), the obtained results are in part ambiguous or contradictory---despite much progress---even at the conceptual level. In particular, there remains ambiguity as to the very basic questions:
(1) At what energy does the many-body localization-delocalization crossover in a quantum dot take place?
(2) What is the width of the crossover? Does the crossover become a sharp transition, for proper scaling, in the limit of a large dot?
(3) What is the behavior of observables in the localized and delocalized phases, and in the critical region?

Our goal is to provide answers to the above questions within a unifying analytical framework. Among the observables, we particularly focus on the many-body level statistics and on temporal relaxation of various types of excited many-body states. Although we concentrate here on localization in Fock space of a finite fermion system, we expect that the solution of the quantum dot problem will be essential for answering the open questions concerning many-body localization in spatially extended systems.

\section{Formulation of the problem}
\label{s1}

We begin by briefly reminding the reader about the formulation of the problem and the relevant parameters. We consider a disordered quantum dot with the mean single-particle level spacing $\Delta$ (the average single-particle density of states is given by $1/\Delta$) and the dimensionless conductance $g\gg 1$. The system is isolated from the ``external world'': the conductance $g$ characterizes the interior of the quantum dot. The Thouless energy $E_T$ is given by the product $g \Delta$. The Hamiltonian of the system reads
\begin{equation}
H=\sum \varepsilon_i c^\dagger_i c_i + \frac{1}{2}\sum V_{ijkl} c^\dagger_i c^\dagger_j c_k c_l~,
\label{Hamiltonian}
\end{equation}
written in the basis of single-particle orbitals with random energies $\varepsilon_i$. The single-particle states are coupled by a two-particle interaction whose matrix elements $V_{ijkl}$ are taken to be random quantities with zero mean (the Hartree-Fock terms are assumed to be accommodated in the first term in $H$). For (internally) screened Coulomb interaction, the root-mean-square matrix element~\cite{blanter96,aleiner01}
\be
V\sim\Delta/g
\label{V}
\ee
for energies of the interacting single-particle states all within an energy band of width $E_T$. The matrix elements for couplings outside this band are strongly suppressed and play no role in what follows.

Being closed, the system is characterized by a discrete spectrum (of both single-particle and many-particle states). That is, in contrast to spatially extended systems, the perturbation theory in powers of the interaction cannot diverge and no broadening of the spectra is generated in higher-order resummations. The coupling between single-particle states only leads to their mutual hybridization, i.e., to a spreading of an excitation in Fock space. It is the ``depth'' of this spreading that is the subject of the problem. In terms of observables, the spreading can be probed, e.g., in the tunneling spectrum by slightly opening the quantum dot through a tunnel coupling to the lead(s).

More specifically, in the absence of interaction, the many-body eigenstates are Slater determinants built out of the eigenstates of the disordered single-particle Hamiltonian. These Slater determinants form the basis of Fock space. The interaction-induced hybridization of the basis states raises the question on how the exact eigenstates in the presence of interaction are spread over the basis states. This question can be viewed as one about the degree of localization of the many-body eigenstates in Fock space, which has much in common with the problem of Anderson localization of noninteracting particles in real space. Two limiting cases are: (i) strong localization, when an exact state almost coincides with one of the noninteracting states, with a small admixture of a few neighboring Fock-space basis states, and (ii) strong delocalization, when an exact state is spread over all the basis states that have energies within the energy band given by the golden rule. One of the main problems here, which we address below, is that of the parameters characterizing the crossover (or the transition, provided the appropriate scaling limit exists) between these two extremes with varying excitation energy $E$.

For zero temperature, we parametrize a many-body excitation by its total energy $E$ and its single-particle content. For given $E$, depending on the single-particle content, two limits of prime interest are a ``hot electron'' excitation and a ``typical'' many-body excitation. In the first case, a single electron with energy $E$ is excited above the many-body ground state. This is the problem that was considered in Ref.~\cite{altshuler97}. In the second case, the excitation energy $E$ is divided between
\be
N_E \sim (E/\Delta)^{1/2}
\label{N_E}
\ee
quasiparticles whose characteristic single-particle energy determines the ``effective temperature'' of the system
\be
T_E\sim (E\Delta)^{1/2}~.
\label{T_E}
\ee
As a characteristic example, one can think of a basis state with $N_E$ quasiparticles. In the typical many-body excitation, the total energy $E \sim N_E T_E$ scales quadratically with the number of excited quasiparticles: $E\sim \Delta N_E^2$. Note that the number $N_\text{mb}$ of excited many-body states with energy $E$ grows exponentially with $N_E$, namely $\ln N_\text{mb} \sim N_E$, i.e.,
\be
\ln N_\text{mb}\sim (E/\Delta)^{1/2}~.
\label{N_mb}
\ee

Below, we study the problem of relaxation in Fock space by focusing on three types of excitations with energy $E\gg\Delta$:
(i) a single hot electron ``on top'' of the exact zero-temperature ground state;
(ii) a ``typical'' many-body excitation in the form of a basis state consisting of $N_E$ single-particle states;
(iii) an extra single-particle excitation with the energy of the order of $T_E$ on top of the exact eigenstate
characterized by the effective temperature $T_E$.
Note that problem (iii) has particularly much in common with the problem of charge and/or energy spreading (and, correspondingly, finite-temperature localization) in real space~\cite{gornyi05,basko06}.

The remainder of the paper is organized as follows. In Sec.~\ref{s2}, we investigate how far each of the initial states described above extends in Fock space in the long-time limit. This will allow us, in particular, to establish a ``phase boundary'' between the localized and delocalized phases in the sense of the Anderson transition in Fock space. In Sec.~\ref{sec:statistics}, we show that the localization-delocalization crossover (which, in particular, manifests itself in the change between the Poisson and Wigner-Dyson spectral statistics for many-body energy levels) becomes a sharp transition in the limit of large $g$ and analyze the critical behavior of the level statistics. In Sec.~\ref{s5}, we complement the long-time-limit analysis of Sec.~\ref{s2} by a discussion of the temporal evolution of the initial states.
Finally, in Sec.~\ref{s6}, we compare our results with previous numerical findings and discuss the connection
of the delocalization transition in a quantum dot to the spatial many-body localization.

\section{Spreading of an excitation in Fock space and the localization threshold}
\label{s2}

\subsection{Golden rule}
\label{subs:GR}

Let us start by considering the problem of relaxation of a single-particle excitation with energy $E\gg\Delta$ above the ground state at zero temperature. The elementary process for this kind of relaxation is a decay of the initial single-particle state into a three-particle state (two electrons and one hole) by exciting an electron-hole pair. The characteristic level spacing of the final (three-particle) states is given by
\be
\Delta_3(E)\sim\Delta^3/E^2\ll\Delta~.
\label{delta3}
\ee
This is because there are typically of the order of $E/\Delta$ possibilities to choose an electron and move
it to one of about $E/\Delta$ empty single-particle states to create an electron-hole pair with energy below $E$.

In Ref.~\cite{sivan94}, SIA performed a golden-rule calculation of the decay rate $\Gamma (E)$ of the single-particle excitation
and obtained, for energies below the Thouless energy,
\begin{equation}
\Gamma(E)\sim\frac{V^2}{\Delta_3(E)}\sim\Delta\left(\frac{E}{E_{T}}\right)^2~,\quad E\alt E_T~.
\label{GammaSIA}
\end{equation}
For higher energies,
Ref.~\cite{sivan94} obtained
\be
\Gamma(E)\sim\Delta \left(\frac{E}{E_T}\right)^{3/2}~,\quad E\agt E_T~.
\label{GammaSIAa}
\ee

\subsection{AGKL}
\label{subs:AGKL}

As was pointed out by AGKL~\cite{altshuler97}, the golden-rule analysis is invalidated by Anderson localization in Fock space at sufficiently small $E\alt\epsilon^{**}$ (in the notation of Ref.~\cite{altshuler97}). AGKL argued that the problem can be mapped onto the one of localization on the Bethe lattice~\cite{abouchacra73}, i.e., a tree with a constant branching number $K$.
In this way, they found
\begin{equation}
\epsilon^{**}\sim E_{1/2}=(g/\ln g)^{1/2}\Delta~.
\label{AGKL-epsilon**}
\end{equation}
For $E\ll\epsilon^{**}$, the exact many-body states are close to the noninteracting
Slater determinants and the initial single-particle state only weakly
hybridizes with other states. The energy $\epsilon^{**}$ was obtained in Ref.~\cite{altshuler97} from
the Bethe-lattice localization threshold~\cite{abouchacra73}
\begin{equation}
W/V\sim K\ln K~,
\label{Z=Zc}
\end{equation}
where $W$ is the width of the (uniform) distribution of on-site energies
and $V$ is the hopping matrix element. In the quantum-dot problem,
\begin{equation}
W=K\Delta_3(E)~,
\label{W-AGKL}
\end{equation}
and $V$ is given by Eq.~(\ref{V}).
Solving Eqs.~(\ref{Z=Zc}) and (\ref{W-AGKL}) for $E$ yields $E\sim \Delta (g/\ln K)^{1/2}$, which reduces to Eq.~(\ref{AGKL-epsilon**})
for $K$ being a power-law function~\cite{altshuler97} of $g$. The same result holds
if one takes into account that the characteristic value of $K$ is~\cite{note1}
\be
K\sim (E/\Delta)^2~.
\label{K}
\ee

The overall picture proposed by AGKL is as follows. For $E>\epsilon^{**}$, the initial single-particle state is well connected to (a subset of) distant many-body states that are constructed out of the maximum number $N_E$ [Eq.~(\ref{N_E})] of single-particle states (``delocalized regime''). However, in a logarithmically wide interval
$\epsilon^{**}<E<\epsilon^*$, where
\begin{equation}
\epsilon^*\sim g^{1/2}\Delta~,
\label{AGKL-epsilon*}
\end{equation}
the single-particle tunneling density of states (DOS), given by the overlap matrix elements between the single-particle states and the exact eigenstates,
experiences strong fluctuations (in the sense that the sharp resonances in the DOS---recall that the system is finite---that group together do not show a regular pattern for the envelope of a group). The energy scale $\epsilon^*$ corresponds to the condition
\begin{equation}
V\sim \Delta_3(E)~,
\label{V=Delta3}
\end{equation}
which implies that a single-particle state at energy $E$ is strongly hybridized with available three-particle states.
Equivalently, Eq.~(\ref{V=Delta3}) is rewritten in terms of the SIA golden-rule decay rate $\Gamma(E)$ [Eq.~(\ref{GammaSIA})] as
\begin{equation}
\Gamma(E) \sim \Delta_3(E)~,
\label{Gamma=Delta3}
\end{equation}

One way to interpret Eq.~(\ref{Gamma=Delta3}) is that a self-consistent broadening of the single-particle levels (if one employs the notion of self-consistency in the spirit of the golden-rule calculation) at this energy is of the order of the relevant (three-particle) level spacing. The difference in the logarithmic factor between $\epsilon^*$ and $\epsilon^{**}$ is traced back to the contribution to the hybridization of rare resonances between the sites in  Fock space~\cite{supplementary}. For $E\gg\epsilon^*$, the peaks in the single-particle DOS (which are clusters of the ``partial peaks'' that describe multiple channels for the hybridization of a given single-particle excitation) acquire a regular Lorentzian envelope whose width is given by the golden-rule decay rate.

The AGKL results for the local DOS (in particular, its fluctuations) on a Bethe lattice with large connectivity $K$
were corroborated by an analysis using the supersymmetry technique~\cite{mirlin97}.
On the other hand, the mapping of the quantum-dot problem to the Bethe lattice model
that was put forward by AGKL turns out to be not fully correct,
as we are going to discuss in what follows.

\subsection{Relaxation of a hot electron revisited}
\label{subs:beyond}

The main approximation made by AGKL in Ref.~\cite{altshuler97} was the replacement of the actual graph in Fock space by the Bethe lattice (a tree with a constant branching number). As already briefly mentioned in Sec.~\ref{s0}, this approximation is insufficient for determining the chaotization threshold (in the finite system) for the following two reasons.

(i) The branching number is not constant: it decreases with increasing number of steps in the hybridization process.
This introduces a factorial in the generation number $n$ suppression of the coupling between the initial single-particle
state and the final many-particle states~\cite{jacquod97,mirlin97,silvestrov98}.
This is because higher-order matrix elements for transitions between distant states
in Fock space are dominated by the contributions of intermediate (virtual) states that approximately conserve the total energy along the paths
in Fock space. This implies that the effective phase space for higher-order processes is reduced, since the total
energy $E$ is redistributed among many single-particle states. As a result of the energy partitioning, the level
spacing for elementary processes at generation $n$ is determined by $E/n$ rather than $E$, see Eq.~(\ref{Delta-n}) below.

\begin{figure}
\centerline{\includegraphics[width=9cm]{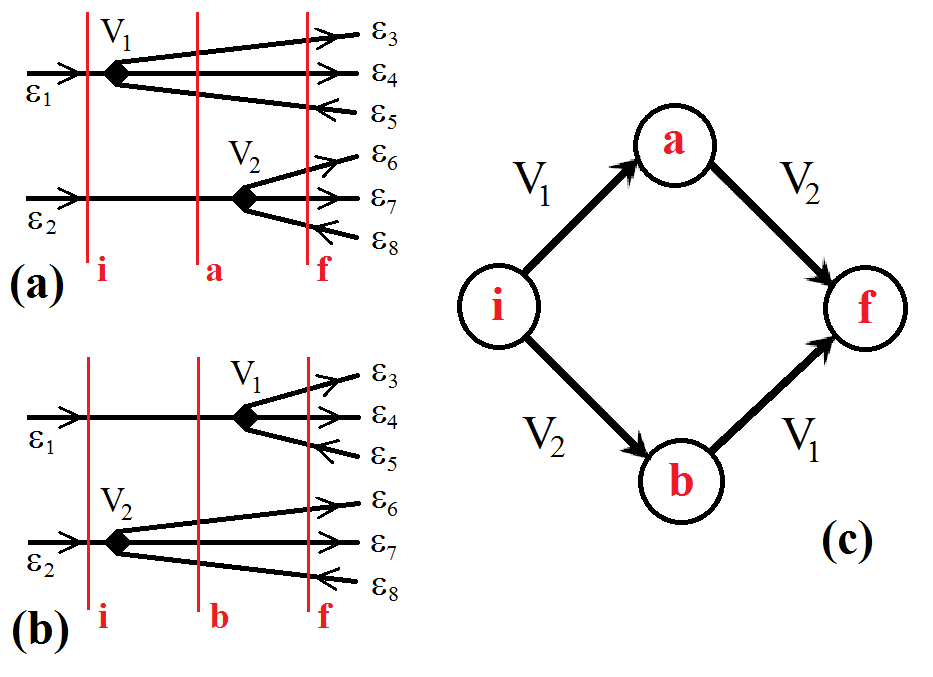}}
\caption{Illustration of a decay process of $p=2$ quasiparticles into 6 quasiparticles.
(a),(b) Diagrams for Schr\"odinger perturbation theory with two different virtual states a and b, respectively.
Red vertical lines indicate the sequence of many-body states. (c) Corresponding interfering paths in Fock space. }
\label{fig:Feynman}
\end{figure}

(ii) The topology of the actual graph in Fock space is more complex than that of the Bethe lattice:
the paths connecting the initial and final states are strongly correlated, see Fig.~\ref{fig:Feynman}. The correlated paths correspond to different time orderings of the same interaction event in the Schr\"odinger perturbation theory while yielding a single Feynman diagram, as was pointed out in Refs.~\cite{silvestrov97} and \cite{silvestrov98}, and later discussed in the context of spatial many-body localization in Ref.~\cite{basko06} and, in particularly great detail, Ref.~\cite{ros15}. The quantum interference between the paths, again, factorially in $n$ reduces the effective coupling between distant states in Fock space.

To illustrate the origin of the suppression factor (ii), we consider a $p$-particle state (which may be the initial one or arise at certain order of the perturbative expansion), as shown in Fig.~\ref{fig:Feynman} for $p=2$. Each of the $p$ particles forming this state can decay into a three-particle state. The elementary interaction events can be time-ordered in $p!$ ways, thus yielding $p!$ different paths on the effective lattice in Fock space. The contributions of the different paths are, however, strongly correlated with each other. First, the product of the
elementary matrix elements, connecting the neighboring states along the paths that are only different in the time ordering, is the same for each
of the paths. Second, the on-site energies on the Fock-lattice are not independent from each other.
 In fact, the correlated terms largely cancel, yielding a single (with the factorial accuracy; the actual number of terms has been analyzed in Ref.~\cite{ros15}) term rather than $p!$. This leads to the $1/p!$ factor in comparison to what one would get on a lattice with the same coordination number but without correlations, see Eq.~(\ref{coupling-n}) below.

%%%%%%%%%%%%%%%%%%%%%%%%%%%%%%%%%%%%%%%%%%%%%%%%%%%%%%%%%%%
%%%%%%%%%%%%%%%%%%%%%%%%%%%%%%%%%%%%%%%%%%%%%%%%%%%%%%%%%%

In the example presented in Fig.~\ref{fig:Feynman}, there are two distinct paths in Fock space that connect the same initial (i) and final (f) states but differ by the intermediate many-body states (a and b). Note that the decay processes of quasiparticles 1 and 2 are completely independent of each other, while the paths in Fock space, Fig.~\ref{fig:Feynman}(c), interfere with each other.
The energies corresponding to the cross-sections in diagrams \ref{fig:Feynman}(a) and \ref{fig:Feynman}(b) read:
\begin{eqnarray}
\varepsilon_i\!&=&\!\varepsilon_1+\varepsilon_2,\quad \varepsilon_f=\varepsilon_3+\varepsilon_4+\varepsilon_6+\varepsilon_7-\varepsilon_5-\varepsilon_8,
\nonumber
\\
\varepsilon_a\!&=&\!\varepsilon_2+\varepsilon_3+\varepsilon_4-\varepsilon_5,\quad
\varepsilon_b=\varepsilon_1+\varepsilon_6+\varepsilon_7-\varepsilon_8,
\nonumber
\\
\label{states}
\end{eqnarray}
where $\varepsilon_{1},\ldots,\varepsilon_{7}$ are the energies of single-particle states involved.
The mixing of the initial and final states is described by the second-order term of the Schr\"odinger (stationary) perturbation theory,
\begin{eqnarray}
\frac{V_1V_2}{\varepsilon_f-\varepsilon_i}\left(\frac{1}{\varepsilon_a-\varepsilon_i}+\frac{1}{\varepsilon_b-\varepsilon_i}\right)
&=& \frac{V_1V_2}{(\varepsilon_a-\varepsilon_i)(\varepsilon_b-\varepsilon_i)},
\nonumber\\
\label{V1V2}
\end{eqnarray}
where we have used $\varepsilon_a+\varepsilon_b-2\varepsilon_i=\varepsilon_f-\varepsilon_i$, as it follows from Eq.~(\ref{states}).
Thus we see that instead of two terms for two paths in the second-order correction we have a single term.
This statement can be generalized for an arbitrary order of perturbation theory~\cite{basko06,ros15}.
This can be done, in particular, by employing the following algebraic identity (used in Ref.~\cite{ros15}),
\begin{equation}
\sum_{\text{permutations}}
\!\!\frac{1}{\mathcal{E}_1({\mathcal{E}_1\!+\!\mathcal{E}_2})\!\ldots\!({\mathcal{E}_1\!+\!\mathcal{E}_2\!+\!\ldots\!+\!\mathcal{E}_n})}
=\frac{1}{\mathcal{E}_1\mathcal{E}_2\ldots \mathcal{E}_n},
\label{permutations}
\end{equation}
which reduces a sum over $n!$ permutations of the energy labels to a single term.
Here the notation $\mathcal{E}_i$ is used for the energy difference
at the $i$th interaction vertex.

As mentioned above, the cancellation of $n!$ can also be seen from Feynman diagrams.
Specifically, as was shown in Refs. \cite{gornyi05,ros15}, the number of topologically distinct connected
Feynman diagrams, $N_d$, contains no factorials and grows only exponentially with increasing $n$.
The number of different arrangements of the pole integration within a single diagram, $N_c$, also scales
exponentially~\cite{ros15}, so that
the total number of terms at $n$th order (given by the product $N_dN_c$) does not contain any factorial factors $n!$.

Let us now determine the chaotization threshold by taking into account the above factors (i) and (ii).
For the ease of presentation, we first neglect the logarithmic ($\ln g$) factors [which have the same origin as the $\ln K$ factor on the Bethe lattice, Eq.~(\ref{Z=Zc})] that will be restored in the end of the calculation.
Furthermore, for this estimate, it is sufficient to only consider the scattering processes that lead to the increase of the number of excited quasiparticles (i.e., to neglect the recombination processes). For the corresponding paths in Fock space, the generation number $n$ is equal to the order of the perturbation theory. In any state of generation $n$, there are, then, $2n+1$ quasiparticles (which by itself is similar to the definition of the generation number in Ref.~\cite{altshuler97}), each with a typical energy of the order of $E/n$, created without loops in Fock space at any step.

Starting with the initial single-particle state with energy $E$, the characteristic level spacing $\Delta^{(1)}(E)$ for the directly coupled states of generation 1 is $\Delta_3(E)$ [Eq.~(\ref{delta3})]. As $n$ increases, for a given state of generation $n$, the characteristic level spacing for the directly coupled states of generation $n+1$ grows linearly in $n$:
\begin{equation}
\Delta^{(n)}(E)\sim\frac{1}{n}\,\Delta_3\!\left(E/ n\right)\sim n\,\frac{\Delta^3}{E^2}~.
\label{Delta-n}
\end{equation}
The factor $1/n$ in front of $\Delta_3(E/n)$ in Eq.~(\ref{Delta-n}) reflects that each of the $2n+1$ already excited quasiparticles can excite an extra electron-hole pair. The argument of $\Delta_3(E/n)$ is the characteristic energy $E/n$ of the quasiparticles. The coupling of the initial state of energy $E$ to the states of the $n$th generation is controlled by the parameter
\begin{equation}
\eta_n(E)=\frac{1}{n!}\frac{V}{\Delta^{(1)}}\,\ldots\,\frac{V}{\Delta^{(n)}}\,\sim\frac{1}{(n!)^2}\left(\frac{E^2}{g\Delta^2}\right)^n~.
\label{coupling-n}
\end{equation}
In the $1/(n!)^2$ dependence of $\eta_n$ on $n$, one factor $1/n!$ (written explicitly in the first equality) comes
from the destructive quantum interference of multiple paths connecting two sites of the lattice [suppression factor (ii) at the beginning of Sec.~\ref{subs:beyond}]. The other factorial results from the decrease, as $n$ increases, of the branching number on the lattice in Fock space [suppression factor (i)], as encoded in $\Delta^{(n)}$, Eq.~(\ref{Delta-n}).

The result for the coupling constant (\ref{coupling-n}) can also be obtained
diagrammatically~\cite{gornyi05,ros15}.  The factor $(n!)^{-2}$, which distinguishes Eq.~(24)
from the analogous quantity on the Bethe lattice, comes then from the density of $(2n+1)$-particle
states that are directly accessible from the initial state in the hybridization spreading described by
a single diagram.

Many-particle states of generation $n$ are hybridized with the initial single-particle state of energy $E$ if the coupling $\eta_n(E)\agt 1$, which gives the maximum generation $n(E)$ hybridized for given $E$:
\begin{equation}
n(E)\sim \frac{E}{g^{1/2}\Delta}~.
\label{En}
\end{equation}
The characteristic maximum generation number for given $E$ is $N_E$ from Eq.~(\ref{N_E}) (higher orders of the perturbative expansion necessarily involve recombination processes). Equating $n(E)$ and $N_E$ and solving the resulting equation for $E$, we thus find the threshold energy $E_c$ at which the single-particle state can hybridize with many-particle states of all generations. Restoring the resonant logarithmic factors, the coupling $\eta_n(E)$ in Eq.~(\ref{coupling-n}) should be multiplied by $(\ln g)^n$, which gives
\begin{equation}
E_c \sim E_1=\frac{g}{\ln g} \Delta~.
\label{Ec}
\end{equation}

Note that repeating the same steps without the factor $(1/n!)^2$ in Eq.~(\ref{coupling-n}) would lead to the AGKL result $E_c\sim\epsilon^{**}$ with $\epsilon^{**}$ from Eq.~(\ref{AGKL-epsilon**}).
In view of the suppression factors (i) and (ii) this energy scale, however, does not correspond
to the mixing of the initial state with many-particle states up to the largest possible generation $n\sim N_E$ and thus does not correspond to the many-body delocalisation transition. The energy $\epsilon^{**}$  marks the characteristic scale at which the hybridization only starts to develop: the number of generations that are hybridized at $E\sim\epsilon^{**}$ is of the order of unity. As $E$ is increased, the number of hybridized generations $n\sim E/\epsilon^{**}$ grows linearly in $E$, yielding Eq.~(\ref{Ec}) for the energy at which the hybridization extends through the entire Fock space. We thus arrive at the important conclusion that the delocalization energy is, in fact, much larger, by a factor $(g/\ln g)^{1/2}$, than the one proposed by AGKL.

Finally, note that, compared to AGKL, Refs.~\cite{jacquod97} and \cite{mirlin97} took into account the decreasing branching number for the Fock-space lattice [factor (i)] but missed the effect of correlations between the paths on the lattice, thus keeping only one factor $1/n!$, out of necessary two, in Eq.~(\ref{coupling-n}). This led, erroneously, to the conclusion that the threshold energy is given by $E_{2/3}$ [Eq.~(\ref{e2})], which is parametrically smaller than $E_c$ in Eq.~(\ref{Ec}). The energy (\ref{e2}) still provides a characteristic scale for the problem: as we discuss in Sec.~\ref{subs:typical}, Eq.~(\ref{e2}) gives a characteristic energy at which a ``typical'' state begins to hybridize substantially with adjacent (in Fock space) states.

\subsection{Relaxation of a typical many-body state}
\label{subs:typical}

Consider now the initial state that we termed in Sec.~\ref{s1} a ``typical'' many-body state, i.e.,  a basis state in which its energy $E$ is distributed among $N_E$ [Eq.~(\ref{N_E})] single-particle excitations. The overlap of the initial state with any many-body eigenstate is assumed to be small. All single-particle states within the energy band of width $T_E$ [Eq.~(\ref{T_E})] are available for the hybridization. The level spacing of the many-body states directly coupled to the initial state is now given by Eq.~(\ref{Delta-n}) with $n\sim N_E$:
\begin{equation}
\Delta^{(N_E)}(E)\sim \Delta \left(\frac{\Delta}{E}\right)^{3/2}~.
\label{DeltaNtyp}
\end{equation}
The condition $\Delta^{(N_E)}(E)\sim V$ yields the energy at which a typical many-body state is hybridized with its Fock-space neighbors.
This energy is of the order of $E_{2/3}$ [Eq.~(\ref{e2})].
It is worth noting that this scale does not contain a logarithmic factor, in contrast to Eq.~(\ref{Ec}).
In the limit $g\to \infty$, it is smaller than the delocalization threshold (\ref{Ec}).

The coupling of a typical state to those of generation $n$ is parametrized by
\be
\eta^{\rm typ}_n(E)\sim \frac{1}{n!}\left[\frac{V}{\Delta^{(N_E)}(E)}\right]^n\sim \left(\frac{E^{3/2}}{ng\Delta^{3/2}}\right)^n~,
\label{coupling_typical}
\ee
instead of Eq.~(\ref{coupling-n}). The difference compared to Eq.~(\ref{coupling-n}) is that the relevant level spacing of the $n$th generation does not grow with $n$ but is independent of $n$ and given by Eq.~(\ref{DeltaNtyp}). For given $E$, the number of hybridized generations, obtained from the condition $\eta^{\rm typ}_n(E)\agt 1$, is then
\be
n(E)\sim \frac{N_E^3}{g}\sim \frac{E^{3/2}}{g\Delta^{3/2}}~.
\label{n_Etyp}
\ee
Substituting $N_E$ for $n(E)$ in this condition, we recover
the delocalization threshold in the form of Eq.~(\ref{Ec}), but without the logarithmic factor in the denominator.
In fact, once the energy exceeds the one given by Eq.~(\ref{Ec}), processes of consecutive decay of quasiparticles (represented by connected Feynman diagrams, cf. Ref.~\cite{silvestrov98}) become important, ensuring the delocalization (see also Sec.~\ref{subs:single} below). This restores the $\ln g$ factor in the denominator of the delocalization threshold Eq.~(\ref{Ec}) \cite{supplementary,FOOTNOTE-1}.

\subsection{Relaxation of a single particle on top of a thermal state}
\label{subs:single}

As another important example, consider the hybridization of an initial state constructed by adding a single-particle excitation to a background ``thermal state''. We use the term ``thermal state'' here to describe a many-body eigenstate with energy $E$ and the single-particle content that is characterized by the effective temperature $T_E$ in the sense of Eqs.~(\ref{N_E}) and (\ref{T_E}). Alternatively, one can think of a mixed state with the equilibrium density matrix characterized by the temperature $T_E$. The latter setting is particularly close to the one used in Refs.~\cite{gornyi05} and \cite{basko06} for spatially extended systems.

The coupling of the initial single-particle excitation to many-body states of generation $n$ is now parametrized by
\be
\eta_n^T(E)\sim\frac{1}{n!}\left[\frac{nV}{\Delta_3(T_E)}\right]^n\sim\left(\frac{E}{g\Delta}\right)^n~.
\label{coupling-T}
\ee
In contrast to Eq.~(\ref{Delta-n}), the energy $E$ is now not partitioned between higher generations, i.e., the characteristic energy of three-particle excitations at each step of the hybridization process does not depend on $n$ and is given by $T_E$. Note that, in contrast to Eqs.~(\ref{coupling-n}) and (\ref{coupling_typical}), the $n!$ factors cancel out in the coupling constant (\ref{coupling-T}), which makes $\eta_n^T(E)$ particularly similar to the coupling on the Bethe lattice. Restoring the $\ln g$ factor, the delocalization energy, corresponding to $\eta_n^T(E)\sim 1$, is again given by Eq.~(\ref{Ec}). Note that, substituting $E_c$ for $E$ in Eq.~(\ref{T_E}), one obtains $T_c\sim (g/\ln g)^{1/2} \Delta$ for the value of the threshold
\textit{temperature} of the thermal state.

\section{Level statistics: Transition between Poisson and Wigner-Dyson}
\label{sec:statistics}

In Sec.~\ref{s2}, the various types of the initial state all give the same delocalization energy $E_c$ [Eq.~(\ref{Ec})].
For energies below the threshold,
the hybridization of the initial state stops well below the maximum generation $N_E$. This clearly corresponds to the Poisson
statistics of many-particle levels. For $E>E_c$, the hybridization goes up to the maximum generation.
The characteristic number of hybridized quasiparticles in the vicinity of the delocalization transition follows from Eq.~(\ref{N_E})
with $E_c$ substituted for $E$:
\be
N_{E_c}\sim (g/\ln g)^{1/2}~.
\label{Nc}
\ee
The total number of many-body states available at the transition is given by Eq.~(\ref{N_mb}) with $E\sim E_c$, which can be rewritten in terms of the many-body level spacing at the critical point $\Delta_{{\rm mb},c}$:
\be
\ln (\Delta_{{\rm mb},c}/\Delta)\sim -(g/\ln g)^{1/2}~.
\label{DeltaMB}
\ee
Although, as discussed in Sec.~\ref{subs:beyond}, the lattice in Fock space (of a finite, in real space, correlated electron system) is essentially different from the Bethe lattice, the exponentially small level spacing in Eq.~(\ref{DeltaMB}) indicates similarity between the Fock-space problem and the noninteracting Bethe-lattice problem (and related models, such as the sparse random matrix model~\cite{sparse}). Specifically, the similarity is in that the effective spatial dimensionality of both lattices is infinite. What is crucial for this conclusion is that the Hilbert space for either of the lattices grows with the number of sites exponentially, rather than as a power law, which would be the case for finite dimensionality.

At energies above the Thouless energy, $E\gg E_T$, the golden rule applies~\cite{sivan94} and all states are well mixed within the SIA golden-rule width, Eq.~(\ref{GammaSIAa}). (We remind the reader that $E_T$ exceeds $E_c$ by a factor of the order of $\ln g$.) In the intermediate range
$E_c \alt E \alt E_T$ fluctuations of delocalized eigenstates are strong, in analogy with the corresponding regime on the Bethe lattice, see Refs.~\cite{altshuler97,mirlin97}.

Let us now discuss the level statistics at and above $E_c$.  Exactly at the critical point, $E=E_c$, a noninteracting disordered system exhibits a critical statistics intermediate between Poisson and Wigner-Dyson statistics. This critical statistics depends on spatial dimensionality $d$, approaching the Poisson statistics with increasing $d$.  Since the present model corresponds to an infinite effective dimensionality, we expect that the many-body
level statistics, which is Poissonian in the localized phase ($E < E_c$) independently of dimensionality, should remain Poissonian also at the
critical point ($E = E_c$).

An important question is then at what $E>E_c$ a crossover to the Wigner-Dyson spectral statistics occurs.
For $E>E_c$, the hybridization extends over the whole Fock space, the size of which is limited by $N_{\rm mb}$ [Eq.~(\ref{N_mb})].
Experience with disordered single-particle systems in real space tells us that the crossover between two types of spectral statistics
is controlled by the ratio of the effective Thouless energy and the level spacing. Therefore, the Wigner-Dyson statistics will be
applicable at the many-body level-spacing scale $\Delta_{{\rm mb},c}$ under the condition that the effective many-body
Thouless energy $E_{T,{\rm mb}}$ (the characteristic inverse time of diffusion through the whole many-body Fock space) exceeds $\Delta_{{\rm mb},c}$.

To find $E_{T,{\rm mb}}(E)$ in the delocalized phase near the transition, we need to know the critical behavior of the Fock-space diffusion coefficient $D_{\rm mb}(E)$. For the Bethe lattice, the continuous vanishing of the diffusion coefficient at the critical point was analyzed in much detail in Refs.~\cite{zirnbauer86,efetov87,mirlin91}. The most characteristic feature of the transition on the Bethe lattice is that while the correlation length diverges at the transition as a power law of the distance to the critical point in units of $W/V$ [Eq.~(\ref{Z=Zc})] (similar to the conventional continuous transition), the diffusion coefficient vanishes exponentially fast. That the diffusion coefficient vanishes faster than any power law is a direct consequence of the infinite dimensionality of the Bethe lattice~\cite{remark_Dmb}. Since this property of the Bethe lattice is shared by the lattice in Fock space [as discussed right below Eq.~(\ref{DeltaMB})], we expect $D_{\rm mb}(E)$ to behave as a function of $E-E_c$ for $E\to E_c$ exponentially as well:
\begin{equation}
\ln D_{\rm mb}(E)\sim -C(g)\left(\frac{E_c}{E-E_c}\right)^{\kappa}~.
\label{DMB}
\end{equation}

If there were a direct mapping on the Bethe lattice in the problem of relaxation in a quantum dot, the results of Refs.~\cite{zirnbauer86,efetov87,mirlin91} for the diffusion coefficient on the Bethe lattice would translate, as it was discussed in Ref.~\cite{mirlin97}, into $\kappa=1/2$ and $C(g)\sim\ln g$ in Eq.~(\ref{DMB}). The actual absence of the direct mapping leaves the question about the exponent $\kappa$ open. As far as the function $C(g)$ is concerned, it cannot be faster than logarithmic, because $D_{\rm mb}(E)$ for $E\gg E_T$ is given by the golden-rule formula and $E_T$ is only $\ln g$ times larger than $E_c$. Although the exact shape of $D_{\rm mb}(E)$ in Eq.~(\ref{DMB}) is beyond the scope of this paper, Eq.~(\ref{DMB}) allows us to answer the question about the nature of the crossover between the Poisson and Wigner-Dyson regimes.

Clearly, the Fock-space Thouless energy $E_{T,{\rm mb}}(E)$ has the same exponential scaling behavior as the diffusion coefficient
$D_{\rm mb}(E)$,  Eq.~(\ref{DMB}).
For $g\gg 1$, when solving the equation $E_{T,{\rm mb}}(E)\sim\Delta_{{\rm mb},c}$ for $E$, it suffices, with logarithmic accuracy, to compare the arguments of the exponential functions in Eqs.~(\ref{DeltaMB}) and (\ref{DMB}). The solution of this equation gives the characteristic energy $E_{\rm WD}$ above which the level statistics is close to the Wigner-Dyson form. The result reads:
\begin{equation}
E_{\rm WD}-E_c\sim E_c\left[\,\frac{C^2(g)\ln g}{g}\,\right]^{\frac{1}{2\kappa}}~.
\label{crit-width}
\end{equation}
The relative width of the critical region for the level statistics, $(E_{\rm WD}-E_c)/E_c$, is seen to vanish in a power-law fashion in the limit $g\to \infty$. In this sense, there is a sharp transition between the localized (Poisson) and delocalized (Wigner-Dyson) phases in the large-$g$ limit.

\section{Temporal decay of excitations}
\label{s5}

In Sec.~\ref{s2}, we analyzed how far the various initial states that we chose as characteristic examples spread in Fock space in the limit of large time. A related important question is how this spreading develops in time. One of the quantities that characterize the time evolution is the return probability
\begin{equation}
P(t)=\left| \langle i| e^{-i Ht}| i \rangle \right|^2~,
\end{equation}
i.e., the probability to find the system in the initial state $|i\rangle$ in time $t$. In the context of relaxation in Fock space for a finite correlated system, this quantity was considered earlier in Refs.~\cite{flambaum01} and \cite{silvestrov01}.

Below, we analyze $P(t)$ for the three types of the initial state that were introduced in Sec.~\ref{s1}, with the final spreading of the states in Sec.~\ref{s1} being characterized by the limiting value of $P(t\to\infty)$.
Similar to Sec.~\ref{s1}, the overlap of the initial states with any of the exact many-body eigenstates is assumed to be small. It is important to note that, for a given realization of disorder in the isolated system, $P(t)$ shows periodic revivals~\cite{rivas02,vasseur15}. These are, however, suppressed upon disorder averaging (which we assume to be performed below).

At this point, it is worth mentioning the connection between the return probability and the IPR.
The disorder-averaged limiting value of $P(t\to \infty)$ can be cast in the form
\begin{equation}
I_i = P(t\to \infty)=\sum_\alpha \left|\langle \alpha |i \rangle\right|^4,
\label{IPR-1}
\end{equation}
where the summation goes over all exact many-body eigenstates $|\alpha\rangle$.
The quantity $I_i$, which characterizes the expansion of a given state $|i\rangle$ over eigenstates (see, e.g., Refs. \cite{silvestrov97,leyronas99,rivas02}), is frequently referred to as the IPR of the state $|i\rangle$. This quantity should be distinguished from the (more conventional)
IPR of exact many-body eigenstates defined as
\begin{equation}
I_\alpha=\sum_m \left|\langle m | \alpha \rangle\right|^4.
\label{IPR-2}
\end{equation}
This IPR describes the content of the eigenstate $|\alpha\rangle$ in terms of the basis states $|m\rangle$, see, e.g., Refs.~\cite{silvestrov98,leyronas00},
and is analogous to the definition of the IPR of eigenfunctions $\psi_\alpha(\bf x)$ of a noninteracting problem, $I_\alpha = \int d^d x |\psi_\alpha^4(\bf x)|$. If one chooses a typical basis state [state (ii) introduced in Sec.~\ref{s1}] as the initial state $|i\rangle$, its IPR $I_i$, Eq.~(\ref{IPR-1}), will essentially coincide with Eq.~(\ref{IPR-2}).

As a starting point in the analysis of the return probability $P(t)$, let us consider a noninteracting system on the Bethe lattice with hopping matrix element $V$ and level spacing of neighboring states $\Delta_3$, well above the transition (i.e., in the golden-rule regime). Let us assume that the initial state is localized on a single site. To find the law $P(t)$ of the temporal decay of this state, one can relate the time $t$ with the dominant generation number $n$ at this time.
Since a random walk on the Bethe lattice with a large coordination number
has quasi-ballistic character, one has a linear relation
\begin{equation}
n\sim \Gamma t,
\label{ntBethe}
\end{equation}
with the golden-rule rate $\Gamma \sim V^2/\Delta_3$.
Further, the return probability can be estimated as the inverse number of coupled sites $\mathcal{N}_n$
which grows exponentially with $n$,
\begin{equation}
-\ln P(t) \sim n.
\label{Ptn}
\end{equation}
[For simplicity, we drop logarithmic prefactors in the relation (\ref{Ptn}).]
Thus, we find for the Bethe lattice
\begin{equation}
-\ln P(t) \sim \Gamma t,
\label{PtBethe}
\end{equation}
i.e., the exponential decay.

In what follows, we extend this analysis to the evolution in Fock space of the many-body system (quantum dot). As was explained in Sec.~\ref{subs:beyond}, there are two key aspects in which the Fock-space graph differs from the Bethe lattice: (i) variation of the coordination number with generation and (ii) presence of multiple correlated paths connecting the initial and final states. We thus have to analyze how these properties affect the temporal decay law
in a quantum dot.

\subsection{Hot electron}
\label{s5.1}

We begin by analyzing the time evolution of a single-particle excitation created with energy $E$ on top of the many-body ground state, which is the problem considered by Silvestrov~\cite{silvestrov01}. We assume that $E\gg g^{1/2}\Delta$, so that the hybridization spreads over many generations [Eq.~(\ref{En})] also in the insulating phase. It is convenient, similarly to the Bethe-lattice case discussed above, to describe the decay process in terms of a time-dependent generation number $n(t)$ and the corresponding scattering rate $\Gamma_n$  (cf. Ref.~\cite{flambaum01}),
\be
\partial_tn = \Gamma_n(t).
\label{Gamma_n_def}
\ee
For the hot-electron excitation, $\Gamma_n$ has the form
\begin{equation}
\Gamma_n\sim n\Gamma(E/n)~,
\label{dndt}
\end{equation}
where $\Gamma (E/n)$, given by Eq.~(\ref{GammaSIA}) or (\ref{GammaSIAa}), is the characteristic decay rate for one of $2n+1$ quasiparticles of generation $n$, the characteristic energy of each of which is $E/n$, through the creation of an electron-hole pair. For $E\alt E_T$, we have
\be
\Gamma_n\sim \frac{E^2}{ng^2\Delta}~.
\label{Gamma_n}
\ee

Note that Eq.~(\ref{Gamma_n}) can be equivalently obtained as $\Gamma_n\sim V^2/\Delta^{(n)}(E)$ with $\Delta^{(n)}(E)$ from Eq.~(\ref{Delta-n}). The factor $1/n$ in Eq.~(\ref{Gamma_n}) reflects the $1/n$ decrease of the coordination number of the Fock-space lattice, i.e., equivalently, the linear-in-$n$ increase of the many-body level spacing in Eq.~(\ref{Delta-n}). This should be contrasted with the Bethe-lattice case, where the corresponding rate is independent of $n$.

Importantly, no additional $n$ dependent factor similar to that leading to $n!$ in the $n$th-order coupling in Eq.~(\ref{coupling-n}) appears in Eq.~(\ref{Gamma_n}). That is, the $n!$ suppression of the coupling (\ref{coupling-n}) that is due to the compensation of differently time-ordered terms in a single Feynman diagram for the scattering event does not show up explicitly as an $n$ dependent factor in $\Gamma_n(t)$. To see why, it is instructive to discuss the following auxiliary problem, similar to the one discussed in the beginning of Sec.~\ref{subs:beyond}.

Consider, as the initial state, one of the basis states in Fock space with energy $E$ partitioned between $2p+1\gg 1$ quasiparticles. The characteristic level spacing for the states with $2p+3$ quasiparticles, directly connected to the initial state, is given by $\Delta^{(p)}\sim p \Delta^3/E^2$ [Eq.~(\ref{Delta-n})]. On the other hand, the characteristic level spacing for the three-particle states directly connected to a given single-particle state is $\Delta_3(E/p)$. Of interest to us is the interval of $V$ [Eq.~(\ref{V})]
\be
\Delta^{(p)}(E)\ll V\ll \Delta_3(E/p)~,
\label{Ep}
\ee
which exists for $p\gg 1$ (and can be rewritten as $p\ll VE^2/\Delta^3\ll p^2$). The left condition means that a transition to the next-generation state is possible, while the right condition means that individual quasiparticles typically do not decay, so that the transition can only occur due to the excitation of electron-hole pairs by a small fraction of quasiparticles.

The scattering amplitude $a_{\rm if}(t)$ for the decay of a single-particle state $|{\rm i}\rangle$ into a three-particle state $|{\rm f}\rangle$ in time $t$ obeys the two-level formula with the off-diagonal matrix element $V/2$:
\begin{equation}
\left|a_{\rm if}(t)\right|^2=\frac{V^2}{V^2+\omega_{\rm if}^2}\sin^2\frac{\sqrt{V^2+\omega_{\rm if}^2}\ t}{2}
\label{two-level}
\end{equation}
[we take here the root-mean-square matrix element $V$ from Eq.~(\ref{V}) as a characteristic one], which for $t\alt 1/\max\{V,|\omega_{\rm if}|\}$ yields
\begin{equation}
\left|a_{\rm if}(t)\right|^2\sim  V^2t^2~,
\label{V2t2}
\end{equation}
independently of the relation between $V$ and the energy difference of the initial and final states $|\omega_{\rm if}|$.

For $t\ll 1/\Delta_3(E/p)$, a given single-particle state $|i\rangle$ is directly coupled to (of the order of) $1/\Delta_3(E/p)t\gg 1$ three-particle states $|{\rm f}\rangle$. Summing up $|a_{\rm if}(t)|^2$ from Eq.~(\ref{V2t2}) over $|{\rm f}\rangle$ for given $|{\rm i}\rangle$,
\be
\sum_{{\rm f:\,\,|\omega_{\rm if}|}\alt 1/t}|a_{\rm if}(t)|^2\sim V^2t^2\times\frac{1}{\Delta_3(E/p)t}~,
\ee
we reproduce the golden-rule decay probability $P_d(t)$ (for one single-particle state) growing linearly in $t$ as
\be
P_d(t)\sim \Gamma(E/p)t~,
\ee
with the single-particle decay rate $\Gamma(E/p)$ from Eq.~(\ref{GammaSIA}). For $t\ll 1/\Delta_3(E/p)$, the probability of finding the system in the initial $p$-particle state factorizes as $P(t)=\left[1-P_d(t)\right]^p$ and reads
\begin{equation}
P(t)\sim e^{-p \Gamma(E/p) t}\sim\exp\left(-\frac{E^2 t}{pg^2\Delta}\right)~.
\label{retprob}
\end{equation}

For $t\gg 1/\Delta_3(E/p)$, there is typically no three-particle state within the energy band of width $1/t$ around a given single-particle level. The state with $2p+3$ quasiparticles is then connected to the initial state through rare resonant couplings of
\begin{equation}
p_a(t)\sim \frac{p}{\Delta_3(E/p)t}
\end{equation}
``active'' quasiparticles, each of which is coupled to its own three-particle ``partner state'' (with $|\omega|\sim 1/t$) according to Eq.~(\ref{V2t2}). Although the decay mechanism for $t\gg 1/\Delta_3(E/p)$ is essentially different from that for shorter times, $P(t)$ for $t\ll 1/V$ is given by the same expression in terms of $\Gamma(E/p)$ as in Eq.~(\ref{retprob}):
\begin{equation}
P(t)\sim \left(1-V^2 t^2\right)^{p_a(t)}\sim e^{-p\Gamma(E/p)t}~.
\label{retprob1}
\end{equation}

The exponential decay in Eq.~(\ref{retprob1}) saturates, as follows from Eq.~(\ref{two-level}), at $t\sim 1/V$. For larger $t$, of the order of $pV/\Delta_3(E/p)\sim V/\Delta^{(p)}(E)$ out of $2p+1$ single-particle states remain strongly hybridized with their three-particle partners, so that $P(t)$ for $t\to\infty$ reads
\begin{equation}
P(\infty)\sim\exp\left[-\frac{pV}{\Delta_3(E/p)}\right]\sim\exp\left(-\frac{E^2}{pg\Delta^2}\right)~.
\end{equation}
The argument of the exponential function in $P(\infty)$ gives the maximum number of hybridized generations
\be
n(E)\sim pV/\Delta_3(E/p)
\label{n_max}
\ee
for $E$ and $p$ obeying the condition (\ref{Ep}).

Thus, the total decay rate for the ($2p+1$)-particle state in Eqs.~(\ref{retprob}) and (\ref{retprob1}) is (simply) given by the sum $pV^2/\Delta_3(E/p)$ of the golden-rule decay rates for the quasiparticles forming the many-body state, or, equivalently, by the many-body golden-rule decay rate $V^2/\Delta^{(p)}(E)$. What the transparent example teaches us is that, for the many-body decay theory in the $t$ representation, there appear no specific factors reflecting the destructive interference of the differently time-ordered scattering events. This should be contrasted with the emergence of these factors in the couplings in Eq.~(\ref{coupling-n}). The fact of the matter is the correlations between the scattering amplitudes with different time-ordering are implicitly accounted for in the above calculation of $P(t)$. If it were not for the correlations, one might erroneously conclude, judging by the large parameter $V/\Delta^{(p)}(E)\gg 1$ [Eq.~(\ref{Ep})] which does not depend on $n$, that the system is in the delocalized phase. The parameter that controls the hybridization is, however, $V/n\Delta^{(p)}(E)$ [as is explicit in Eq.~(\ref{coupling-n})], with the additional factor $1/n$ reflecting the correlations. As a consequence of that, the spreading stops at $n\sim \Delta^{(p)}(E)/V$, which is precisely what we found above [Eq.~(\ref{n_max})] in the calculation in the $t$ representation.

After this digression, we return to the problem of the hot-electron decay.
Solving Eq.~(\ref{Gamma_n_def}) with $\Gamma_n$ from Eq.~(\ref{Gamma_n}), we find
\begin{equation}
n^2(t)\sim\frac{E^2}{g^2\Delta}t~,
\label{n2t}
\end{equation}
and hence, employing the relation between $P(t)$ and $n$ given by Eq.~(\ref{Ptn}) [this relation stems from the effectively infinite dimensionality of the quantum-dot problem, similar to the Bethe lattice in this respect, as discussed in Sec.~\ref{sec:statistics}],
\begin{equation}
-\ln P(t) \sim n(t)\sim [\,\Gamma(E) t\,]^{1/2}~.
\label{lnPt}
\end{equation}

We thus see that the return probability for a hot electron
in the quantum-dot problem bears a certain similarity to the relaxation on the Bethe
lattice. Importantly, in both cases, the quantum-mechanical return probability can be described
by means of a rate equation for the generation number. However, the results for the Bethe lattice and the quantum dot
[Eqs.~(\ref{PtBethe}) and (\ref{lnPt}), respectively] differ from each other because of the different structures of
the lattices. Specifically, the suppression factor (i) that is caused, as discussed in the beginning of Sec.~\ref{subs:beyond},
by the energy partitioning among excited quasiparticles, leads to the $n$ dependence of the rate $\Gamma_n$
for a hot electron. This, in turn, translates into a stretched-exponential rather than a simple exponential decay.
Further, the correlations (destructive interference in Fock space), giving rise to the suppression factor (ii) in the stationary perturbation theory of Sec.~\ref{subs:beyond}, reveal themselves in the saturation of $P(t)$ in the quantum-dot problem.

The stretched-exponential decay of the hot-electron state in Eq.~(\ref{lnPt}) is in agreement with the result of Ref.~\cite{silvestrov01} (up to a logarithmic factor discarded in the above analysis of the temporal decay). In the localized phase, this behavior of $P(t)$ takes place until the generation number reaches the maximum value $n(E)$ given by Eq.~(\ref{En}). Substituting $n(E)$ for $n$ in Eq.~(\ref{n2t}), we obtain the saturation time $t(E)\sim g^2n^2(E)\Delta/E^2$, i.e.,
\begin{equation}
t(E)\sim g/\Delta~.
\label{tmax}
\end{equation}
Note that this time scale is independent of $E$.
Substituting Eqs.~(\ref{tmax}) and (\ref{GammaSIA}) in Eq.~(\ref{lnPt}),
we obtain the saturation value of the return probability,
\begin{equation}
-\ln P(\infty) \sim \frac{E}{g^{1/2}\Delta}.
\label{Psat-hot}
\end{equation}

\begin{figure}
\centerline{\includegraphics[width=8cm]{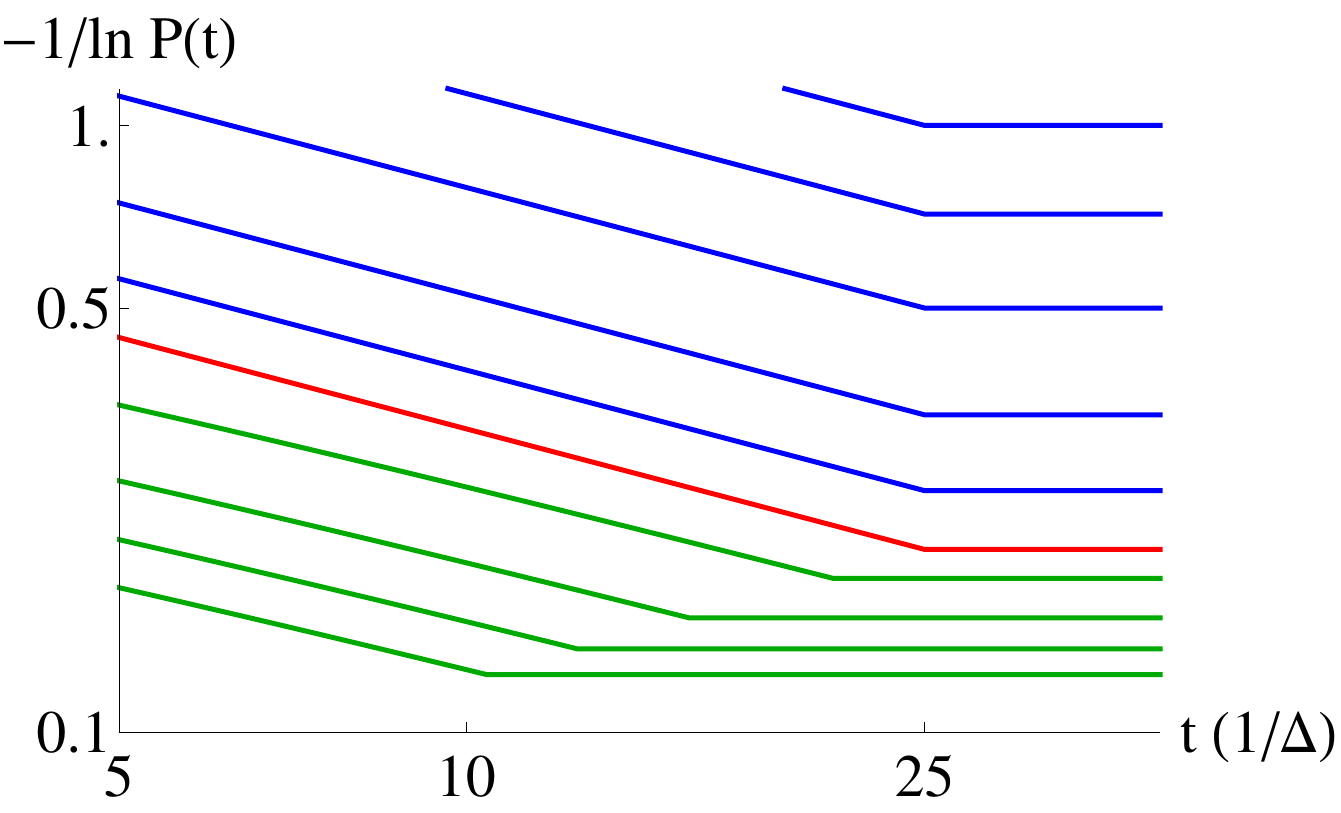}}
\caption{
Schematic illustration of the saturation crossover in the return probability $P(t)$ on the log-log scale
across the delocalization transition
for a hot electron with $E/\Delta = 5,\ 7,\ 10,\ 15,\ 20,$\ $25,\ 30,\ 40,\ 50,\ 60$ (from top to bottom) and $g=25$. Blue lines correspond to the localized regime, green lines to the delocalized regime, and the red line to the localization threshold.  The slope of all lines corresponds to the stretched exponential decay, Eq.~(\ref{lnPt}).
The saturation time is given by Eq.~(\ref{tmax}) in the localized regime and by Eq.~(\ref{t_E_typ}) in the delocalized regime.
 The saturation value of $-\ln  P(T)$ is given by Eq.~(\ref{En}) in the localized regime and by Eq.~(\ref{N_E}) in the delocalized regime.}
\label{fig:Pt}
\end{figure}

In the delocalized phase, the decay proceeds in four steps. As long as the characteristic energy of excited quasiparticles $E/n$ is larger than $E_T$, the rate $\Gamma(E/n)$ in Eq.~(\ref{dndt}) is given by Eq.~(\ref{GammaSIAa}), yielding $(-\ln P)\sim n$ of the form
\begin{equation}
-\ln P(t)\sim \frac{E}{g^{2/3}\Delta^{1/3}}t^{2/3}~, \quad t\alt t_0\sim\frac{1}{g^{1/2}\Delta}~.
\label{nSIA}
\end{equation}
In terms of $n(t)$ [Eq.~(\ref{n2t})], $\ln P(t)$ scales as $t^{2/3}$ for $n\alt n(t_0)\sim E/g\Delta$. For $t\agt t_0$, the elementary scattering rate is given by Eq.~(\ref{GammaSIA}) and the dependence of $P(t)$ crosses over into
\begin{equation}
-\ln P(t)\sim\frac{E}{g \Delta} [\,1+\Delta(t-t_0)\,]^{1/2}~,\quad t\agt t_0~.
\label{Ptt0}
\end{equation}
For $t_0\alt t\alt 1/\Delta$, $P(t)$ obeys
\be
-\ln\frac{P(t)}{P(t_0)}\sim \frac{E}{g}t~.
\label{linear-decay}
\ee
For $t\gg 1/\Delta$, Eq.~(\ref{Ptt0}) becomes Eq.~(\ref{lnPt}). As $t$ increases, the decay in the delocalized phase eventually saturates at
\be
t(E)\sim g^2/E~,
\label{t_E_typ}
\ee
when $n(t)$ reaches the maximum generation number (\ref{N_E}). The behavior of $P(t)$ is illustrated in Fig.~\ref{fig:Pt} for energies below (blue lines) and above (green lines) the localization threshold for $g=25$.

\subsection{Typical many-body state}
\label{s5.2}

For the ``typical'' many-body state with energy $E$ (Sec.~\ref{subs:typical}), the decay rate in the $n$th generation is independent of $n$ and given by
\begin{equation}
\Gamma_n\sim \frac{V^2}{\Delta^{(N_E)}(E)}~,
\label{GammaGR}
\end{equation}
with $\Delta^{(N_E)}(E)$ from Eq.~(\ref{DeltaNtyp}). In contrast to the $t^{1/2}$ scaling of $n(t)$ in Eq.~(\ref{n2t}), $n(t)$ from Eqs.~(\ref{Gamma_n_def}) and (\ref{GammaGR}) changes now linearly with $t$:
\be
n(t)\sim \frac{E^{3/2}}{g^2\Delta^{1/2}}t~,
\label{n-typical}
\ee
which means the ``conventional'' exponential decay
\begin{equation}
-\ln P(t)\sim \frac{E^{3/2}}{g^2\Delta^{1/2}}t~.
\label{exp-decay}
\end{equation}

%%%%%%%%%%%%%%%%%%%%%%%%%%%%%%%%%%%%%%%%%%%%%%%%%%%%%%%%%%%
%%%%%%%%%%%%%%%%%%%%%%%%%%%%%%%%%%%%%%%%%%%%%%%%%%%%%%%%%%

Again, in the localized regime, Eq.~(\ref{exp-decay}) is valid until the localization-induced saturation sets in. Substituting the hybridization limit $n(E)$ [Eq.~(\ref{n_Etyp})] for $n(t)$ in Eq.~(\ref{n-typical}), we find that the saturation time $t(E)$ is given by the same formula (\ref{tmax}) as for the hot-electron initial state. This yields
\begin{equation}
-\ln P(\infty) \sim \frac{1}{g}\left(\frac{E}{\Delta}\right)^{3/2}.
\label{Psat-typ}
\end{equation}
In the delocalized regime, the decay is saturated at the time given by Eq.~(\ref{t_E_typ}) when $n$ reaches the maximum possible generation number $N_E$ given by Eq.(\ref{N_E}).

\subsection{Single particle on top of a thermal state}
\label{s5.3}

Finally, we consider the third type of the initial state that results from introducing a quasiparticle with energy
of the order of $T_E$ on top of a representative thermal eigenstate characterized by the temperature $T_E$,
see Sec.~\ref{subs:single}. The decay rate is now given by
\be
\label{gamma-thermal}
\Gamma_n \sim \frac{nV^2}{\Delta_3(T_E)}\sim\frac{nT_E^2}{g^2\Delta}~.
\ee
Equation (\ref{Gamma_n_def}) with $\Gamma_n$ from Eq.~(\ref{gamma-thermal}) gives an exponential growth of $n(t)$ with increasing $t$:
\be
\label{n-thermal}
\ln n(t) \sim \frac{T_E^2}{g^2\Delta} t~,
\ee
and thus a double-exponential decay of $P(t)$,
\begin{equation}
\ln\,[\,-\ln P(t)\,]\sim \frac{T_E^2}{g^2\Delta} t~.
\label{exp-decay-thermal}
\end{equation}
These results are applicable in the delocalized phase, for $n(t)$ smaller than the maximum generation number of the order of $T_E/\Delta$. Contrary to the results of Secs.~\ref{s5.1} and \ref{s5.2}, Eqs.~(\ref{n-thermal}) and (\ref{exp-decay-thermal}) do not have any region of applicability in the localized phase, $T_E<T_c$, because then the matrix element $V \ll \Delta_3(E)$ is smaller than the relevant level spacing already in the first generation and there occurs no spreading of the single-particle excitation (cf.\ Sec.~\ref{subs:single}).

\subsection{Fock-space diffusion at criticality}
\label{s5.4}

Now let us discuss the time evolution of the generation number $n(t)$ at criticality, i.e., for $E$ in the critical region between $E_c$ and $E_{\rm WD}$ (Sec.~\ref{sec:statistics}). For the initial state from Sec.~\ref{s5.3}, this consideration will then be valid for the whole temporal evolution.  For the initial state from Sec.~\ref{s5.1} or Sec.~\ref{s5.2}, it will hold for the long-time evolution, i.e., after the saturation of the stretched-exponential (respectively, exponential) decay.

As discussed in Sec.~\ref{sec:statistics}, at criticality the many-body Thouless energy $E_{T,{\rm mb}}$ is of the order of the many-body level spacing $\Delta_{\rm mb}$. The latter satisfies [Eqs.~(\ref{Nc}) and (\ref{DeltaMB})]
\be
\label{spacing-critical}
\ln (\Delta_{\rm mb}/\Delta) \sim -N_{E_c}~.
\ee
Substituting here $E_{T,{\rm mb}}$ for $\Delta_{\rm mb}$, recalling that $E_{T,{\rm mb}}$ is the inverse time of diffusion to the boundary of Fock space, and using the self-similarity of the system at the critical point, we find the ``diffusion law'' at criticality:
\be
\label{crit-diff}
n(t)\sim\ln t~.
\ee
The generation number grows with time according to Eq.~(\ref{crit-diff}) until $t$ reaches $1/E_{T,{\rm mb}}$ [i.e., Eq.~(\ref{crit-diff}) is valid for all $n\alt N_{E_c}$].

Recall that, in the problem of a single quasiparticle added to the thermal state, $n(t)$ grows with $t$ exponentially fast in the delocalized phase [\,$\ln n(t)\propto t$, Eq.~(\ref{n-thermal})]. The logarithmic dependence of $n(t)$ at criticality [Eq.~(\ref{crit-diff})]---in the same problem---is another manifestation of the infinite dimensionality of Fock space (the exponential growth of its volume with the distance $n$). In noninteracting problems of a finite dimensionality, the diffusion laws, both in the delocalized phase and at the critical point, would be of a power-law form.

It is also worth noting that the connection between $P(t)$ and $n(t)$ at criticality should be different from the exponential relation $\ln P(t) \sim -n(t)$ characteristic of the decay processes describable by the ``modified golden rule'' with the $n$-dependent rate $\Gamma_n$ [Eq.~(\ref{dndt})]. This is because the critical states are ``sparse''. In the noninteracting Anderson-transition problem, the sparsity is characterized by multifractality~\cite{evers08}. In the limit of an infinite dimensionality, multifractality takes its extreme form and the IPR remains finite (as in the localized phase) at the critical point (as is exemplified by the Bethe-lattice model~\cite{mirlin91} and the sparse random-matrix model~\cite{sparse}). We expect a similar behavior at criticality in the present problem in view of its effective infinite dimensionality. That is, in the critical regime, we expect $P(t)$ to stay constant as $t$ increases despite the growth of $n(t)$ [Eq.~(\ref{crit-diff})].

\section{Summary and discussion}
\label{s6}

\subsection{Summary}

To summarize, we have reanalyzed the problem of localization in Fock space of interacting electrons in a quantum dot. Our key results are as follows:

(1) We have shown that the delocalization threshold for the eigenstates in Fock space is given by Eq.~(\ref{e3}).

(2) The localization-delocalization transition can be efficiently probed by the statistics of many-body energy levels which crosses over from the Poisson  to Wigner-Dyson statistics. We have estimated the relative width of the crossover and found that it vanishes in the limit of a large dimensionless conductance $g$ (or, equivalently, the limit of weak interaction), see Eq.~(\ref{crit-width}). That is, we have found a sharp many-body Anderson localization transition in this limit.

(3) To determine the position of the transition and to characterize the localized and delocalized phases, as well as the critical regime, we have explored relaxation of three types of the initial quantum state. Specifically, we have considered (i) a hot electron above the ground state, (ii) a typical Fock-space basis state with total energy $E$ distributed among $(E/\Delta)^{1/2}$ single-particle states with energies of the order of $(E\Delta)^{1/2}$; (iii) an extra single-particle excitation on top of an exact ``thermal'' state. For each of them, we have studied the long-time limit of the Fock-space spreading as well as the temporal evolution of the decay process (see Fig.~\ref{fig:Pt} for the case of a hot electron). The energies of the initial states above which a substantial spreading takes place scale according to Eqs.~(\ref{e1}), (\ref{e2}), and  (\ref{e3}) for the states (i), (ii), and (iii), respectively. Above these energies, the return probability scales with time according to Eqs.~(\ref{lnPt}), (\ref{exp-decay}), (\ref{exp-decay-thermal}) until saturation. For any type of the initial state, the spreading reaches the highest generation in Fock space for energies above the critical energy $E_c\sim E_1$ which is given by Eq.~(\ref{e3}).

Our work has thus established unambiguously the physical meaning of the energy scales (\ref{e1}), (\ref{e2}), and (\ref{e3}) (see Sec.~\ref{s0} for an overview of their contradictory discussion in the earlier literature).

\subsection{Comparison to numerical results}

Let us compare our results with the existing numerical findings. Among the numerical works, two papers, Refs.~\cite{leyronas00} and \cite{rivas02}, have presented particularly systematic analysis of the problem. Our result for the position of the localization transition, Eq.~(\ref{e3}), is in full agreement with the numerical result of Leyronas, Silvestrov, and Beenakker, Ref.~\cite{leyronas00}, who calculated the Fock-space inverse participation ratio of eigenstates as a function of energy for various values of $g$. Rivas, Mucciolo, and Kamenev, Ref.~\cite{rivas02}, studied the return probability $P(t)$ for a hot-electron state [initial state of type (i) in our work] by means of a numerical solution of self-consistent equations. Our result for the energy at which the spreading begins, Eq.~(\ref{e1}), is in agreement with their findings. Further, they studied the infinite-time limit of the return probability as a function of energy and found a crossover in its behavior around the energy (\ref{e3}), which is again consistent with our result for the position of the localization transition. However, Rivas, Mucciolo, and Kamenev found an intermediate region between the localized and the golden-rule behavior which becomes broader with increasing $g$. They interpreted this as an evidence in favor of a smooth localization-delocalization crossover rather than a sharp transition. This conclusion~\cite{footnote-RMK} is in conflict with our result,  Eq.~(\ref{crit-width}), which shows that the relative width of the crossover region decreases with increasing $g$, i.e., the transition is sharp in the large-$g$ limit. We offer the following explanation for this apparent discrepancy. One should be careful when using the value of $P(t\to \infty)$ (or the inverse participation ratio) for extracting the position of the transition. Its scaling is given by the localized-phase formula below the critical energy $E_c$, Eq.~(\ref{e3}), and by the golden-rule formula, for energies exceeding $E\sim g\Delta$, which is larger than $E_c$ by a logarithmic factor. In the intermediate regime the states are delocalized but their amplitudes fluctuate strongly, and $-\ln P(t\to \infty)$ is much smaller that one would expect from the golden rule, see the discussion of the analogous region in the Bethe-lattice problem in Refs.~\cite{altshuler97} and \cite{mirlin97}. Thus, it is not surprising that Ref.~\cite{rivas02} observed a broad intermediate regime in the behavior of $P(t\to \infty)$.

An alternative way to explore the transition numerically is to study the many-body level statistics.
Let us emphasize once more that the level statistics shows a sharp crossover from the Poisson to Wigner-Dyson behavior near $E_c$.
Therefore, this approach is expected to be an efficient tool to extract the position of the delocalization transition.
As mentioned in Sec.~\ref{s0}, Jacquod and Shepelyansky reported~\cite{jacquod97} that the crossover between the two types of statistics takes place around the energy scale $E_{2/3}$, Eq.~(\ref{e2}), which parametrically differs from $E_c$ (or $E_\text{WD}$) found in the present work. The discrepancy might presumably be attributed to the small size of the system studied numerically in Ref.~\cite{jacquod97}.

In view of the considerable progress in developing numerical methods for studying many-body systems,
it would be desirable to revisit the quantum-dot problem within computational approaches.
In particular, it would be interesting to verify numerically our theoretical predictions regarding the chaotization
threshold and the critical regime, as well as the time-dependent evolution of various initial states.

\subsection{Comparison to spatial many-body localization}

Before closing the paper, we briefly discuss similarities and differences between the many-body localization transitions in a quantum dot (as analyzed here) and in an extended system with spatially localized states (Refs.~\cite{gornyi05} and \cite{basko06}; see also Sec.~\ref{s0} for the references to subsequent publications).

\subsubsection{Position of the transition.}

The quantum-dot localization-transition energy $E_c$ given by Eq.~(\ref{e3}) can  be translated, with the help of Eq.~(\ref{T_E}), into an effective transition temperature
\be
\label{Tc-quantum_dot}
T_c\sim (E_c\Delta)^{1/2} \sim \Delta(g/\ln g)^{1/2}~.
\ee
This result arises in a particularly transparent way when one considers the decay of a single-particle excitation on top of a thermal state [our initial state (iii), see Sec.~\ref{subs:single}], which is a direct counterpart of the analysis in Refs.~\cite{gornyi05} and \cite{basko06}. In full analogy with Eq.~(\ref{e4}), the result (\ref{Tc-quantum_dot}) can be obtained (up to a logarithmic factor) by comparing the matrix element with the level spacing of directly coupled states in this setting. We remind the reader that $1/g$ is the dimensionless strength of the effective (screened) interaction in the quantum-dot problem [Eq.~(\ref{Tc-quantum_dot})], and thus is a counterpart of $\alpha$ in Eq.~(\ref{e4}).
The difference in powers of $1/g$ and $\alpha$ in two formulas is due to opposite relations between the critical temperature $T_c$ and the single-particle Thouless energy in a quantum dot and in the localization volume of extended systems, respectively.

Note that the result $E_{2/3}$ of Ref. \cite{jacquod97} for the transition energy for a typical state [our setup (ii)]
translates by means of Eq.~(\ref{T_E}) into the effective temperature $T_c \sim g^{1/3}\Delta$.
If the same logic as in Ref. \cite{jacquod97} (comparing the matrix element with the level spacing of directly
coupled typical many-body states) was used, as in Refs. \cite{[BG00], [DLS01]}, for extended systems, this would yield
a power-low vanishing of the transition temperature, $T_c \to 0$, with increasing system size,
at variance with Eq. (\ref{e4}).

\subsubsection{Return probability.}

The return probability $P(t)$ analyzed in Sec.~\ref{s5} for the quantum-dot case
can be studied in extended systems as well. A particularly close analogy holds
for the relaxation of typical (basis) states. The return probability
in the localized phase of extended systems was discussed, in particular, in the context of quantum quenches.
It is worth mentioning that in extended systems
there is a relation between the saturation of $P(t\to \infty)$ and the saturation of the
entanglement entropy $S_\text{ent}(t)$, see Refs.~\cite{bardarson12} and \cite{serbyn13},
though the entropy approaches the saturation value much more slowly.
The saturation values of $S_\text{ent}(t)$ and of the participation ratio [that determines $P(t\to \infty)$]
were found to be proportional to the system size. The underlying mechanism responsible
for such behavior is similar to that governing the golden-rule decay of the initial state
before the saturation of $P(t)$, see Sec.~\ref{s5.1}.
Since questions regarding the entanglement entropy require dividing
the system into two subsystems in real space, it is not straightforward to formulate a
direct analogy in the quantum-dot problem.

\subsubsection{Scaling of the decay rate near the transition.}

On the delocalized side of the transition ($E>E_c$) in the quantum-dot problem, the level width emerges which scales as
\begin{equation}
\Gamma \propto \exp\left[-C(g)\left(\frac{E_c}{E-E_c}\right)^{\kappa}\right].
\label{width-quantum-dot}
\end{equation}
Within this energy window, eigenstates are strongly correlated and the level statistics has the Wigner-Dyson form
(for $E>E_\text{WD}$), see Sec.~\ref{sec:statistics}. (We refer the reader to Refs.~\cite{sparse} and \cite{mirlin97} for the discussion of an analogous scale in the sparse-matrix and Bethe-lattice models.)
Clearly, Eq.~(\ref{width-quantum-dot}) has the same form in terms of temperatures with  $T\sim (E\Delta)^{1/2}$ and $T_c$ given by Eq.~(\ref{Tc-quantum_dot}).
It was argued in Ref.~\cite{gornyi05} that a similar scaling of the quasiparticle decay rate (imaginary part of the self-energy) emerges
close to the transition in the delocalized phase  of a system with spatially localized single-particle states,
\begin{equation}
\Gamma \propto \exp\left[-C(\alpha)\left(\frac{T_c}{T-T_c}\right)^{\kappa}\right].
\label{width-spatially-loc}
\end{equation}
This result with $\kappa=1/2$ was found in Ref.~\cite{gornyi05} from an approximate mapping of the problem to the Bethe-lattice model.
Gopalakrishnan and Nandkishore~\cite{gopalakrishnan14} found the result (\ref{width-spatially-loc}) from a numerical analysis of self-consistent equations. Their fitting of numerical data suggests, however, $\kappa=1/3$.  Monthus and Garel~\cite{monthus10} obtained numerically the critical behavior of the type (\ref{width-spatially-loc}) with $\kappa=1.4$ for a certain renormalized hopping characterizing the decay. Quite generally, the behavior (\ref{width-spatially-loc}) is a consequence of the effective infinite-dimensional character of the phase. Whether the value
$\kappa=1/2$, which follows from the Bethe-lattice approximation, is actually an exact result remains to be clarified by future work. All in all, the critical behavior of the level width in the quantum-dot problem, Eq.~(\ref{width-quantum-dot}), and in systems with spatially localized single-particle states, Eq.~(\ref{width-spatially-loc}), are qualitatively similar. Details, such as the values of $\kappa$ and the scaling of the factors $C(g)$ and $C(\alpha)$, may turn out to be different, though.

\subsubsection{Scaling of the diffusion constant (or the conductivity) near the transition.}

In the quantum-dot problem, the level width (\ref{width-quantum-dot}) yields the ``diffusion coefficient'' (\ref{DMB}) for the evolution in Fock space.
In the simplest scenario, this applies also to the problem with spatially localized single-particle states.
This scenario assumes that the level width $\Gamma$, Eq.~(\ref{width-spatially-loc}),
determines also the spatial diffusion coefficient and thus the conductivity,
\begin{equation}
\sigma \propto \exp\left[-C(\alpha)\left(\frac{T_c}{T-T_c}\right)^{\kappa}\right].
\label{conductivity-spatially-loc}
\end{equation}
This assumption was made in Ref.~\cite{gornyi05} and later in Ref.~\cite{gopalakrishnan14}.
The critical point at $T=T_c$ entails a transition from the Poisson to Wigner-Dyson statistics, in full analogy with the above discussion of the quantum-dot problem. The form of the anomalous diffusion at the critical point $T_c$ can be found by equating the inverse time of diffusion through the system and the many-body level spacing, where the latter satisfies
\begin{equation}
- \ln \Delta_{\rm MB}\sim L^d
\label{DMB-d}
\end{equation}
for a $d$-dimensional system of spatial size $L$. Requiring
$$
\int_0^{1/\Delta_{\rm MB}} dt D(t) \sim L^2,
$$
we get the scaling of the time-dependent diffusion coefficient $D(t)$, or, equivalently, of the mean-square displacement $\langle r^2(t)\rangle$ at criticality:
\begin{eqnarray}
D(t) &\sim& \frac{1}{t}(\ln t)^{2/d-1} ,
\label{D-critical-d}  \\
\langle r^2(t)\rangle & \sim& (\ln t)^{2/d}.
\label{r2-critical-d}
\end{eqnarray}
This logarithmically slow spatial diffusion at critically is a counterpart of the critical Fock-space diffusion (\ref{crit-diff}) in the quantum-dot problem. For a related recent result see Ref.~\cite{serbyn15}.  Equation~(\ref{D-critical-d}) implies the following critical scaling of the ac conductivity at low frequencies:
\be
\sigma(\omega) \sim |\omega|(\ln|\omega| )^{2/d-1}.
\label{sigma-critical-d}
\ee

It is worth reiterating that a rigorous foundation for theory of the critical behavior at the MBL transition
in extended systems, especially, on the delocalized side, is still lacking and the above discussion represents the simplest scenario.
Specifically, the following should be mentioned in this context:

(i) The assumption that the level width $\Gamma$ directly determines the effective transport time and thus the diffusion coefficient (\ref{conductivity-spatially-loc}) is by no means self-evident and requires verification.
Within the analysis of Ref.~\cite{gornyi05}, the localized phase becomes unstable due to ``ballistic processes'' (those with necklace structure in the terminology of Ref.~\cite{ros15}) that can be mapped onto the Bethe-lattice problem, leading to Eq.~(\ref{width-spatially-loc}) for the level width.
In Ref.~\cite{basko06}, similar processes (termed there self-avoiding paths) were argued to be responsible for the critical behavior on the localized side of the transition.
While leading to the delocalization of certain types of many-body excitations, these processes \textit{by themselves} are not sufficient, however, to establish a finite dc conductivity~\cite{future}. This leaves a possibility for the existence of an (at least) intermediate phase with finite $\Gamma$ but zero dc conductivity (the effective ``transport scattering rate'' in this phase is not given by $\Gamma$).

(ii) The authors of Refs.~\cite{agarwal15,vosk15,ACP15,knap15} argued that Griffiths effects induce an intermediate delocalized phase with subdiffusive charge transport (and thus vanishing dc conductivity).

(iii) Recent numerical results appear to favor the emergence of a subdiffusive delocalized phase in one-dimensional \cite{agarwal15,barlev15,gopalakrishnan15,lerose15} and two-dimensional~\cite{reichmann15} systems.

We relegate a more detailed investigation of these intriguing issues to future work.

\section*{Acknowledgements}

We would like to thank M.~V. Feigel'man, L.~B. Ioffe, V.~E. Kravtsov, I.~V. Protopopov, D.L.~Shepelyansky,
M.~A. Skvortsov, and K.~S. Tikhonov for useful discussions.
We are particularly grateful to P.G.~Silvestrov for numerous insightful conversations and comments on our paper.
This work was supported by Russian Science Foundation under
Grant No.\ 14-42-00044 (I.V.G. and A.D.M.).

\newpage

\widetext

\begin{center}
\textbf{SUPPLEMENTAL MATERIAL}
\end{center}

\renewcommand{\thepage}{S\arabic{page}}
\renewcommand{\theequation}{S\arabic{equation}}
\renewcommand{\thefigure}{S\arabic{figure}}

\setcounter{page}{1}
\setcounter{section}{0}
\setcounter{equation}{0}
\setcounter{figure}{0}

\vspace*{0.25cm}

\begin{center}
\parbox{16cm}
{\small In Supplemental Material, we discuss the origin of the logarithmic factor in the energies $E_c$ and $E_{1/2}$ [Eqs.~(26) and (13) of the main text, respectively]. The former is the delocalization threshold, the latter marks the characteristic energy scale at which a hot-electron state starts to hybridize with neighboring states in Fock space. We also explain the role of correlations between different contributions to the perturbative expansion of the hybridization amplitude. These correlations lead to the absence of a logarithmic factor in the energy $E_{2/3}$ [Eq.~(2)], at which a typical basis state starts to hybridize with its Fock-space neighbors.}
\end{center}

\vspace*{0.75cm}

For the noninteracting model on the Bethe lattice, the logarithmic factor $\ln K$ in the condition for the delocalization threshold [Eq.~(14)] follows from the solution of the corresponding self-consistency equation \cite{abouchacra73,efetov85,zirnbauer86,efetov87,verbaarschot88,mirlin91,mirlin97}. In fact, this logarithm was found already in Anderson's 1958 paper \cite{anderson58}. Below, we discuss the reason for the appearance of a similar logarithm in our problem [in Eqs.~(13) and (26)] by analyzing---to a large extent along the lines of Refs.~\cite{anderson58,altshuler97,ros15}---the condition for the existence of resonances in the hybridization process. We put a particular emphasis on the effect of possible correlations between different contributions to the perturbative expansion of the hybridization amplitude. This allows us to clarify the presence or absence of the logarithmic factor in the characteristic energy scales of the quantum-dot problem.

We start our discussion with a simple exercise in probability theory (we will explain its connection to the localization problem afterwards). Consider a set $\{a_i\}=(a_1, a_2, \ldots, a_K)$ of $K$ random numbers distributed independently over the interval $[0,1]$. Denote $a_\text{min}$ the minimum number $\text{min}\left\{a_i\right\}$ in a given realization of the set $\left\{a_i\right\}$. Introduce the function $\mathcal{F}_a(x)$ as the probability that all numbers of the set
$\left\{a_i\right\}$ are larger than a given number $x$:
\begin{eqnarray}
\mathcal{F}_a(x)&=&\text{Prob}\Big\{\text{all}\ a_i>x\Big\}
=\prod_i \int_x^1\! da_i\, \mathcal{P}(a_i)~,
\label{SEq2}
\end{eqnarray}
where $\mathcal{P}(a_i)$ is the probability density for the distribution of $a_i\in [0,1]$.
Let all $a_i$ be distributed homogeneously over the interval $[0,1]$ with
\begin{equation}
\mathcal{P}(a_i)
=\theta(a_i)\theta(1-a_i)~.
\label{SEq1}
\end{equation}
For the set $\left\{a_i\right\}$ with the probability density (\ref{SEq1}), we have
\begin{equation}
\mathcal{F}_a(x)=\prod_{i=1}^{K} \int_x^1 da_i=(1-x)^K.
\label{SEq3}
\end{equation}
Define the \emph{typical minimum value} $a_\text{min}^\text{typ}$ for the set $\left\{a_i\right\}$ as the value of $x$
such that $\mathcal{F}_a(x)=1/2$:
\begin{equation}
\mathcal{F}_a\!\left(a_\text{min}^\text{typ}\right)=\frac{1}{2}~.
\label{SEq4}
\end{equation}
For the set $\left\{a_i\right\}$ described by Eq.~(\ref{SEq1}), the condition (\ref{SEq4}) yields
\begin{equation}
a_\text{min}^\text{typ}=1-\frac{1}{2^{1/K}}~,
\label{SEq5}
\end{equation}
which for large sets
becomes
\begin{equation}
a_\text{min}^\text{typ}\simeq \frac{1}{K} \ln 2~, \qquad K\gg 1~.
\label{SEq6}
\end{equation}

Now, consider a ``composite set,'' constructed, as the simplest example, of only two uncorrelated random numbers $(a,b)$, each of which is homogeneously distributed over the interval $[0,1]$, and look at the statistical properties of the product $X=ab$.
The probability density $\mathcal{P}(X)$ for the distribution of $X\in [0,1]$ obeys
\begin{equation}
\mathcal{P}(X)=\int_0^1 da \int_0^1\!db\, \delta(X-ab)
=\theta(X)\theta(1-X)\ln\frac{1}{X}~.
\label{SEq7}
\end{equation}
Note the appearance of the logarithm in the distribution of the product $X=ab$ already for the two-dimensional set $(a,b)$. The question we now face is the one of the typical minimum value $X_\text{min}^\text{typ}$, defined similar to Eq.~(\ref{SEq4}), for a composite set $\{X_{ij}\}$ of $K^2$ elements with $i,j=1,\ldots, K$. Our purpose is to show that the $K$ dependence of $X_\text{min}^\text{typ}$ for large $K$ depends in an essential way on precisely how the composite set is formed out of completely {\it uncorrelated} random numbers. Specifically,
we illustrate this point by considering the following three possibilities of generating the set $\{X_{ij}\}$
with $i,j=1,\ldots, K$:
\begin{itemize}
\item[]\ $X^{(A)}:\quad {X_{ij}^{(A)}\!=\!a_{ij}b_{ij}}$\, with $K^2$ elements $a_{ij}$ and $K^2$ elements $b_{ij}$;
\item[]\ $X^{(B)}:\quad {X_{ij}^{(B)}\!=\!a_{i}b_{ij}}$\, with $K$ elements $a_{i}$ and $K^2$ elements $b_{ij}$;
\item[]\ $X^{(C)}:\quad {X_{ij}^{(C)}\!=\!a_{i}b_{j}}$\, with $K$ elements $a_{i}$ and $K$ elements $b_{j}$.
\end{itemize}
In all three sets $X^{(A),(B),(C)}$, there are no correlations of the elements of the subset $a$ with the elements of the subset $b$, each subset
of the random numbers being governed by Eq.~(\ref{SEq1}).

Consider first $X^{(A)}$. For this set, the function $\mathcal{F}_A(x)$ [defined similar to Eq.~(\ref{SEq2})] factorizes into the product of $K^2$ terms:
\begin{equation}
\mathcal{F}_A(x)=\!\prod_{1\leq i,j \leq K} \left[ \int_0^1\! da_{ij}\! \int_0^1\! db_{ij}\,\int_x^1\! dX_{ij}\, \delta(X_{ij}-a_{ij}b_{ij}) \right]
=\left(1-x-x \ln\frac{1}{x}\right)^{K^2}
\stackrel{K\gg 1}{\simeq}
\exp\left(-K^2 x \ln \frac{1}{x}\right)~.
\label{SEq8}
\end{equation}
Equating $\mathcal{F}_A(x)$ and $1/2$ [Eq.~(\ref{SEq4})], we find:
\begin{equation}
X^{(A)}\, : \qquad X^\text{typ}_\text{min}\simeq \frac{\ln 2}{K^2 \ln K^2}, \qquad K\gg 1~.
\label{SEq9}
\end{equation}
Comparing this result with Eq.~(\ref{SEq6}), we see that the typical minimum value for the composite set $X^{(A)}$ is smaller by the logarithmic
factor $\ln K \gg 1$ than the typical minimum value we would obtain for a simple set $\left\{a_i\right\}$ with the same total number of elements $K^2$.

Consider now $X^{(B)}$. The function $\mathcal{F}_B(x)$ differs from the function $\mathcal{F}_A(x)$. This is because, for given $i$, the integration over any $b_{ij}$ produces the same factor $1/a_i$ (altogether, the factor $1/a_i^K$ after $K$ integrations over $b_{ij}$ for given $i$):
\begin{equation}
\mathcal{F}_B(x)=\!\prod_{1\leq i \leq K}
\left\{ \int_0^1\!da_{i}\!\!\prod_{1\leq j \leq K}\left[\int_0^1\!db_{ij} \int_x^1\!\!dX_{ij}\, \delta(X_{ij}-a_{i}b_{ij})\right]\right\}
=\left[ \int_x^1\! da\left(a-x\right)^K a^{-K}\right]^K~.
\label{SEq10}
\end{equation}
Changing the variable of integration $a\to z=x/a$, we obtain, for $K\gg 1$ and $x\ll 1$:
\begin{eqnarray}
\mathcal{F}_B(x)
\!&\simeq&\!\left(x \int_x^\infty\! \frac{dz}{z^2}\, e^{- K z}\right)^K\! =\! \left[e^{-Kx} + Kx\,\text{Ei}(-Kx)\right]^K
\!\simeq\!
\left\{
\begin{array}{ll}
\exp\left[-K^2 x-K\ln(K x)\right],&  \quad 1/K\ll x\ll 1~, \\[0.2cm]
\exp\left(-K^2 x
\ln\dfrac{1}{K x}
\right),&  \quad x\ll 1/K~,
\end{array}
\right.
\end{eqnarray}
where $\text{Ei}(z)$ is the exponential integral function. The function $\mathcal{F}_B(x)$ decays on the scale of $x\sim K^{-2}/\ln K \ll 1/K$, so that,
to find $X_\text{min}^\text{typ}$, only the asymptotics for $x\ll 1/K$ is relevant,
which gives
\begin{equation}
X^{(B)}\, : \qquad X^\text{typ}_\text{min}\simeq \frac{\ln 2}{K^2 \ln K}, \qquad K\gg 1~.
\label{SEq11}
\end{equation}
Equations (\ref{SEq9}) and (\ref{SEq11}) only differ in the replacement $K^2\to K$ in the argument of the logarithm in the denominator in the latter case.
That is, the typical minimum value for the composite set $X^{(B)}$ is again smaller, by the logarithmic factor $\ln K \gg 1$, than
the typical minimum value for a simple set $\left\{a_i\right\}$ with the same total number of elements $K^2$.

Now turn to $X^{(C)}$. For this set, in contrast to $X^{(B)}$, both $a_i$ and $b_j$ for given $i$ and $j$ appear each in the arguments of $K$ delta-functions. It is convenient to integrate out $X_{ij}$ first:
\begin{eqnarray}
\mathcal{F}_C(x)\!\!&=&\!\! \left(\prod_{1\leq i \leq K}\int_0^1\!\!da_i\right)\! \left(\prod_{1\leq i \leq K}\int_0^1\!\!db_j \right) \!
\left[\prod_{1\leq i,j \leq K}\int_x^1\! dX_{ij}\,  \delta(X_{ij}-a_{i}b_{j})\right]
\nonumber
\\
&=&
\int_0^1\!\!da_1\!\ldots\!\int_0^1\!da_K\int_0^1\!\!db_1\!\ldots\!\int_0^1\!db_K \!\!
\prod_{1\leq i,j \leq K}\!\! \theta(a_i b_j-x)~.
\label{s14}
\end{eqnarray}
Next, order the variables, $a_1\leq a_2\leq \ldots a_K$ and $b_1\leq b_2\leq \ldots b_K$, so that the product of the step functions in Eq.~(\ref{s14}) reduces to a single step function for the smallest variables $a_1$ and $b_1$, namely $\theta(a_1b_1-x)$:
\begin{equation}
\mathcal{F}_C(x)=(K!)^2\!\int_0^1\!da_1\!\int_{a_1}^1\!da_2\!\ldots\!\int_{a_{K-1}}^1\!\!\!\!\!da_K
\!
\int_0^1\!db_1\!\int_{b_1}^1\!db_2\!\ldots\!\int_{b_{K-1}}^1\!\!\!\!\!db_K\, \theta(a_1b_1-x)
\!=\!
K\!\int_x^1\!\! da(1-a)^{K-1}\left(1-\frac{x}{a}\right)^K.
\label{SEq12}
\end{equation}
Assume that $K\gg 1$ and $x\ll 1$. The behavior of $\mathcal{F}_C(x)$ changes at $x\sim 1/K^2$ within this interval:
\begin{equation}
\mathcal{F}_C(x) \simeq  \left\{
\begin{array}{ll}
\sqrt{\pi K} x^{1/4} (1-\sqrt{x})^{2K},&  \quad 1/K^2\ll x\ll 1~, \\[0.2cm]
1-K^2 x \ln\dfrac{1}{K^2x},&  \quad x\ll 1/K^2~.
\end{array}
\right.
\label{SEq13}
\end{equation}
Again, we see that the typical minimum value $X_\text{min}^\text{typ}$ is determined by the small-$x$ asymptotics. Equating $\mathcal{F}_C(x)$ and $1/2$, we find
\begin{equation}
X^{(C)}\, : \qquad X^\text{typ}_\text{min}\simeq \frac{1}{c K^2}, \qquad K\gg 1~,
\label{SEq16}
\end{equation}
where $c\sim 1$ is a number which replaces $\ln K\gg 1$ in the denominator, as compared to Eqs.~(\ref{SEq9}) and (\ref{SEq11}).
This difference comes from the correlations between the elements $X_{ij}^{(C)}$. The simplest correlation of this kind is given by the constraint
\begin{equation}
X_{ij}^{(C)}X_{kl}^{(C)}=a_i b_j a_k b_l = X_{il}^{(C)}X_{kj}^{(C)}~.
\end{equation}
Equations~(\ref{SEq13}) and Eqs.~(\ref{SEq16})] can also be straightforwardly obtained by utilizing the relation, valid for $X^{(C)}$, between the minimum values of the composite and simple sets: $\min\{X_{ij}\}=\min\{a_ib_j\}=\min\{a_i\}\min\{b_j\}$.

Generalizing the above results to composite sets constructed out of $n$ subsets with arbitrary $n$, we find
the following recursion relations:
\begin{eqnarray}
X^{(A)}:\, && X^{(A)}_{i_1i_2\ldots i_n}\!=a^{(1)}_{i_1i_2\ldots i_n}a^{(2)}_{i_1i_2\ldots i_n}\ldots a^{(n)}_{i_1i_2\ldots i_n}:
\qquad \mathcal{F}_A^{(n+1)}(x)=\left\{ x\int_x^1 \frac{dy}{y^2}
\left[ \mathcal{F}_A^{(n)}(y)\right]^{1/K^n} \right\}^{K^{n+1}},
\label{Ir}
\\
X^{(B)}:\, && X^{(B)}_{i_1i_2\ldots i_n}\!=a^{(1)}_{i_1}a^{(2)}_{i_1i_2}\ldots a^{(n)}_{i_1i_2\ldots i_n}:
\qquad
\mathcal{F}^{(n+1)}_B(x)=\left[x \int_x^1 \frac{dy}{y^2} \mathcal{F}_B^{(n)}(y)\right]^K,
\nonumber\\
\label{IIr}
\\
X^{(C)}:\, && X^{(C)}_{i_1i_2\ldots i_n}\!=a^{(1)}_{i_1}a^{(2)}_{i_2}\ldots a^{(n)}_{i_n}:
\qquad
\mathcal{F}_C^{(n+1)}(x)=K x \int_x^1 \frac{dy}{y^2} \left(1-\frac{x}{y}\right)^{K-1} \mathcal{F}_C^{(n)}(y).
\label{IIIr}
\end{eqnarray}
Using these relations, we obtain for $K\gg 1$ and $n\gg 1$ the functions $\mathcal{F}^{(n)}_{A,B,C}(x)$ for $x$ around the point at which these functions cross $1/2$ [see details of the calculation in the end of the file; the function $\text{erf}(z)$ in Eq.~(\ref{III}) is the error function]:
\begin{eqnarray}
X^{(A)}:\, && \quad \mathcal{F}_A^{(n)}(x)\simeq \exp \left[\,-\frac{K^n x}{(n-1)!}\left(\ln\frac{1}{x}\right)^{n-1}\,\right]~,
\label{I}
\\
X^{(B)}:\, && \quad
\mathcal{F}_B^{(n)}(x)\simeq \exp\left[\,-\frac{K^n x}{(n-1)!}\left(\ln\frac{1}{Kx}\right)^{n-1}\,\right]~,
\label{II}
\\
X^{(C)}:\, && \quad
\mathcal{F}_C^{(n)}(x)\simeq \frac{1}{2} -
\frac{1}{2}
\text{erf}\left[\sqrt{\frac{3}{n}} \frac{\ln\left(e^{n\gamma} K^n x\right)}{\pi}\right]~.
\label{III}
\end{eqnarray}
These equations are sufficient for finding  $X_\text{min}^\text{typ}$ for all three sets at $n\gg 1$ :
\begin{eqnarray}
X^{(A)}:\, && X^\text{typ}_\text{min} \sim \frac{1}{K^n (\ln K)^{n-1}}~,
\label{Ia}
\\
X^{(B)}:\, && X^\text{typ}_\text{min}\sim \frac{1}{K^n\left(\ln K \right)^{n-1}}~,
\label{IIa}
\\
X^{(C)}:\, && X^\text{typ}_\text{min}\sim \frac{1}{K^n}~.
\label{IIIa}
\end{eqnarray}
Thus, there is no logarithmic factor in the set $X^{(C)}$ for arbitrary $n$, whereas the powers of the logarithm accumulate with increasing $n$ (in the combination $K\ln K$) in the sets $X^{(A)}$ and $X^{(B)}$.

We are now ready to return to the problem of many-body delocalization in a quantum dot. The existence of resonances in the $n$th generation is controlled by the dimensionless parameter $\eta_n$ determined by the maximum value of ratios
$V^n/\mathcal{E}_1\mathcal{E}_2\ldots \mathcal{E}_n$ representing contributions to the perturbation theory (Feynman diagrams) at $n$th order, with $\mathcal{E}_i$ being energy denominators. Thus, we should look for the probability that the minimum value of the set composed of the products of the energy denominators, $\mathcal{E}_1\mathcal{E}_2\ldots \mathcal{E}_n$, is smaller that $V^n$. This is a problem of precisely the type we considered above.
In Sec. III of the main text, we estimated $\eta_n$ for the three types of initial states (i), (ii), and (iii) [see Eqs. (24), (28), and (30)] by scaling analysis that discards possible logarithmic factors of the same nature as discussed in this Supplemental Material. In order to understand whether the logarithms do emerge in the typical maximum values of $\eta_n$, we compare statistics of the energy denominators with statistics of the random composite sets above.

For the initial states of types (i) and (iii), i.e., for the hot-electron state (Sec.~IIIC) and a single-particle excitation on top of the thermal state (Sec.~IIIE), respectively, we have consecutive decay processes. Namely, an electron decays in three quasiparticles, one of the created quasiparticles decays again into three, etc. This corresponds to a composite set of type B here. Therefore, $\eta_n$ from Eq.~(24) and  $\eta_n^T$ from Eq.~(30) acquire an additional factor $(\ln g)^{n-1}$, which, in turn, results in the appearance of the logarithmic factors in the energy $E_{1/2}$ [Eq. (13)], at which the hybridization begins, and in the energy $E_c$ [Eq. (26)] for the many-body delocalization threshold.

It is worth pointing out that it is the set of type B that also corresponds to the transition in the noninteracting problem on the Bethe lattice. Our results for this set are in full agreement with the emergence of the logarithmic factor $\ln K$ in the equation for the localization threshold [Eq. (14)] that follows from the exact solution for this problem.

On the other hand, for a typical basis state [initial state of type (ii), Sec.~IIID], the hybridization first proceeds via an independent decay of particles that form the initial state; on the level of Feynman diagrams, this is described by disconnected diagrams. This situation corresponds to a set of type C. In this case, no additional logarithmic factors arise in Eq.~(28) for the parameter $\eta_n^{\rm typ}$. Correspondingly, no logarithmic factor appears in Eq.~(2) for the energy $E_{2/3}$, at which a typical basis state starts to hybridize with its Fock-space neighbors.

\begin{center}
\textbf{Calculation of $\mathcal{F}_{A,B,C}^{(n)}$ for $n\gg 1$}
\end{center}

Below, we present the derivation of Eqs.~(\ref{I}),~(\ref{II}), and (\ref{III}).
We start with the set of type A that was analyzed in Ref.~\onlinecite{altshuler97}.
Using the recursion relation Eq.~(\ref{Ir}), we find the exact solution for $\mathcal{F}_A^{(n)}(x)$ at arbitrary $0\leq x\leq 1$
and $K$:
\begin{equation}
\mathcal{F}_A^{(n)}(x)=\left[1-x\sum_{m=0}^{n-1}\frac{\ln^m \frac{1}{x}}{m!}\right]^{K^n}
=\left[1-\frac{\Gamma(n,\ln\frac{1}{x})}{(n-1)!}\right]^{K^n},
\label{FAn}
\end{equation}
where $\Gamma(\alpha,z)$ is the incomplete gamma-function. Indeed, substituting Eq.~(\ref{FAn}) in Eq.~(\ref{Ir}),
we reproduce Eq.~(\ref{FAn}) for $n+1$:
\begin{eqnarray}
\mathcal{F}_A^{(n+1)}(x)\!\!&=&\!\!\left[x\int_x^1 \frac{dy}{y^2} \left(1-y\sum_{m=0}^{n-1}
\frac{\ln^m \frac{1}{y}}{m!} \right)\right]^{K^{n+1}}
\!\!=\left(1-x\sum_{m=0}^{n} \frac{\ln^{m}\frac{1}{x}}{m!} \right)^{K^{n+1}}.
\label{Fa-as}
\end{eqnarray}
For $x\ll 1$ and $K\gg 1$, we keep only the $m=n-1$ term in Eq.~(\ref{FAn}), which gives
Eq.~(\ref{I}). Equating then $\mathcal{F}_A^{(n)}(x)$ and $1/2$, we obtain Eq.~(\ref{Ia}).

Turning to the set of type B, for $x\ll x_{n-1}=K^{1-n}\left(\ln K \right)^{2-n}\!,$ we substitute Eq.~(\ref{II})
in Eq.~(\ref{IIr}):
\begin{eqnarray}
\mathcal{F}_B^{(n+1)}(x)&\simeq &\left\{x \int_x^{x_{n-1}} \frac{dy}{y^2}
\exp\left[\,-\frac{K^n y}{(n-1)!}\left(\ln\frac{1}{Ky}\right)^{n-1}\,\right]~ \right\}^K.
\label{FBint}
\end{eqnarray}
Since we are interested in the behavior of $\mathcal{F}_B^{(n+1)}(x)$ around $x_{n+1}\ll x_n$,
we expand the exponential in Eq.~(\ref{FBint}) and obtain
\begin{eqnarray}
\mathcal{F}_B^{(n+1)}(x)&\simeq &\left\{x \int_x^{x_{n-1}} \frac{dy}{y^2}
\left[\,1-\frac{K^n y}{(n-1)!}\left(\ln\frac{1}{Ky}\right)^{n-1}\,\right]~ \right\}^K
\simeq
\left\{1-\frac{K^n x}{n!}\left(\ln\frac{1}{Ky}\right)^{n} \right\}^K
, \quad x\ll\frac{1}{K^{n}\left(\ln K \right)^{n-1}}~,
\nonumber
\\
\label{FB-as}
\end{eqnarray}
which gives Eqs.~(\ref{II}) and (\ref{IIa}).

Finally, we evaluate $\mathcal{F}_C^{n}(x)$. Since, in contrast to the sets $X^{(A)}$ and $X^{(B)}$, $X_\text{min}^\text{typ}$ for the set $X^{(C)}$ does not contain logarithmic factors in the denominator, we present the calculation in more detail in order to point out where the difference comes from. We start with the definition:
\begin{eqnarray}
\mathcal{F}_C^{(n)}(x)\!\!&=&\!\!
\left[\prod_{1\leq i_1 \leq K}\int_0^1\!\!da^{(1)}_{i_1}\right]\! \ldots
\left[\prod_{1\leq i_n \leq K}\int_0^1\!\!da^{(n)}_{i_n} \right] \!
\left[\prod_{1\leq i_1,\ldots i_n \leq K}\int_x^1\! dX_{i_i\ldots i_n}\,
\delta\left(X_{i_i\ldots i_n}-a^{(1)}_{i_1}a^{(2)}_{i_2}\ldots a^{(n)}_{i_n}\right)\right].
\nonumber
\\
\label{FC}
\end{eqnarray}
First, integrate out all $X_{i_i\ldots i_n}$:
\begin{eqnarray}
\mathcal{F}_C^{(n)}(x)\!\!&=&\!\!
\left[\prod_{1\leq i_1 \leq K}\int_0^1\!\!da^{(1)}_{i_1}\right]\! \ldots
\left[\prod_{1\leq i_n \leq K}\int_0^1\!\!da^{(n)}_{i_n} \right] \!
\left[\prod_{1\leq i_1,\ldots i_n \leq K}\,
\theta\left(a^{(1)}_{i_1}a^{(2)}_{i_2}\ldots a^{(n)}_{i_n}-x\right)\right].
\label{FC1}
\end{eqnarray}
Next, similar to Eq.~(\ref{SEq12}), for each $1\leq m\leq n$, we order the
variables, $a^{(m)}_1\leq a^{(m)}_2\leq \ldots \leq  a^{(m)}_K$,
so that the product of the step functions in Eq.~(\ref{FC1}) reduces to a single step function for the
smallest variables $a^{(m)}_1$, and then integrate out all $a^{(m)}_l$ with $l>1$:
\begin{eqnarray}
\mathcal{F}_C^{(n)}(x)\!\!&=&\!\!
K^n\!
\int_0^1\!\!da^{(1)}_{1}\left[1-a^{(1)}_{1}\right]^{K-1}
\!\ldots\!
\int_0^1\!\!da^{(n)}_{1} \left[1-a^{(n)}_{1}\right]^{K-1}
\theta\left(a^{(1)}_1 a^{(2)}_1\ldots a^{(n)}_1 -x\right)
.
\label{FC2}
\end{eqnarray}
Denoting $a^{(m)}_1=z_m$ and using $\theta(z_1z_2\ldots z_n-x)=\theta[\ln(z_1z_2\ldots z_n)-\ln x]$,
we write
\begin{eqnarray}
\mathcal{F}_C^{(n)}(x)\!\!&=&\!\!
K^n\!
\int_0^1\!\!dz_{1}\left(1-z_{1}\right)^{K-1}
\!\ldots\!
\int_0^1\!\!dz_{n} \left(1-z_{n}\right)^{K-1}
\int_{\ln x}^{\infty}\!\!du\ \int_{-\infty}^\infty \frac{dt}{2\pi}\
\exp\left[i t \left(\sum_{j=1}^n\ln z_j-u\right)\right]
.
\label{FC3}
\end{eqnarray}
The integrals over $z_1, \ldots, z_n$ now factorize, each producing the beta-function $\text{B}(1+K,1+it)$:
\begin{eqnarray}
\mathcal{F}_C^{(n)}(x)\!\!&=&\!\!
\int_{-\infty}^\infty\! \frac{dt}{2\pi}\
\int_{\ln x}^\infty\!\!du\ e^{-i u t}
\left[K \int_0^1\!\!dz\left(1-z\right)^{K-1}e^{i t \ln z}\right]^n
\!=\!\int_{-\infty}^\infty\! \frac{dt}{2\pi}\
\int_{\ln x}^\infty\!\!du\ e^{-i u t}
\left[\frac{\Gamma(1+K)\Gamma(1+i t)}{\Gamma(1+K + i t)}\right]^n\!
.
\nonumber
\\
\label{FC4}
\end{eqnarray}
For $x=0$, the integral over $u$ gives $2\pi \delta(t)$, which yields $\mathcal{F}_C(0)=1$, as it should be.
Integrating out $u$ for arbitrary $0\leq x\leq 1$, we get
\begin{eqnarray}
\mathcal{F}_C^{(n)}(x)\!\!&=&\!\!
\frac{1}{2} + \text{Im} \int_{0}^\infty\! \frac{dt}{\pi}\ \frac{e^{-i t \ln x}}{t}
\left[\frac{\Gamma(1+K)\Gamma(1+i t)}{\Gamma(1+K + i t)}\right]^n~.
\label{FC5}
\end{eqnarray}
In order to calculate $\mathcal{F}_C(x)$ for $K\gg 1$, we employ Stirling's approximation (here $\gamma\approx 0.5772$ is the Euler-Mascheroni constant):
\begin{eqnarray}
\left[\frac{\Gamma(1+K)\Gamma(1+i t)}{\Gamma(1+K + i t)}\right]^n\simeq
\left\{
  \begin{array}{ll}
    \exp\left(-i\, n\, t \ln K\, -\,  i\, n\, \gamma\, t -\dfrac{\pi^2}{12}\, n\, t^2\right),\quad & t\ll 1~, \\[0.4cm]
    (2 i \pi t)^{n/2}\exp\left[-i\, n\, t \left(\ln \dfrac{K}{t}+1\right) -\dfrac{\pi}{2}\, n\, t\right],\quad & 1\ll t \ll K~, \\[0.4cm]
    (2\pi K)^{n/2}\exp\left[-n\, K \left(\ln\dfrac{t}{K}+1\right)-i\dfrac{\pi}{2}\, n\, K\right],\quad & K \ll t~.
  \end{array}
\right.
\label{FC6}
\end{eqnarray}

For large $n\gg 1$, the integral in Eq.~(\ref{FC5}) for arbitrary $0<x<1$ converges already for $t\ll 1$, so that only the first asymptotics in Eq.~(\ref{FC6}) matters. The dependence of the function $\mathcal{F}_C(x)$ on $K$ shows up, then, through a single parameter $\left(Ke^\gamma\right)^n x$:
\begin{eqnarray}
\mathcal{F}_C^{(n)}(x)\!\!&\simeq&\!\!
\frac{1}{2} - \int_{0}^\infty\! \frac{dt}{\pi}\sin\left[t\ln(K^n x) + n\gamma t\right] \frac{e^{-\pi^2 n t^2/12}}{t}
\nonumber
\\
&=&\frac{1}{2} -
\frac{1}{2}
\text{erf}\left(\sqrt{\frac{3}{n}} \frac{\ln\left[\left(e^\gamma K\right)^n x)\right]}{\pi}\right)\quad
 \text{for} \ K\gg 1\ \text{and}\ n\gg 1~,
\label{FC7}
\end{eqnarray}
which is Eq.~(\ref{III}).
Here
\begin{equation}
\text{erf}(z)=\frac{2}{\sqrt{\pi}} \int_0^z e^{-t^2}dt \simeq \left\{
                                                               \begin{array}{ll}
                                                                 \dfrac{2z}{\sqrt{\pi}},\quad & |z|\ll 1, \\[0.2cm]
                                                                 \text{sgn}(z)-\dfrac{e^{-z^2}}{\sqrt{\pi} z},\quad & |z|\gg 1
                                                               \end{array}
                                                             \right.
\label{erf}
\end{equation}
is the error function.
This result follows from the central limit theorem for the distribution of independent logarithms of the minimum
values $z_m=a_\text{min}^{(m)}$ for the independent subsets $a^{(m)}$.
We see that, according to Eq.~(\ref{SEq4}),  $X_\text{min}^\text{typ}$ is given by the zero in the argument
of the error function in Eq.~(\ref{FC7}):
\begin{equation}
X_\text{min}^\text{typ}\simeq \frac{1}{\mathcal{C}^n K^n},\quad n\gg 1,
\end{equation}
with $\mathcal{C}=e^\gamma.$


\begin{thebibliography}{99}
\bibitem{disclaimer1} The Anderson transition on the Bethe lattice~\cite{abouchacra73} was studied on a ``granular''~\cite{efetov85,zirnbauer86,efetov87,verbaarschot88}  (many orbitals per site) version and on the tight-binding~\cite{mirlin91} (single orbital) version of the lattice.
\bibitem{note1} While AGKL argued that $K\sim g^3$ in the first generation, it was shown in Ref.~\cite{mirlin97} that the proper value of the coordination number for $n\sim 1$ is $K\sim (E/\Delta)^2$. However, in the condition (\ref{Z=Zc}) with $W$ from Eq.~(\ref{W-AGKL}), $K$ appears only in the argument of the logarithm. The abovementioned difference in $K$ does not affect the conclusion that $\ln K \sim \ln g$ for $E\sim\epsilon^{**}$, up to a factor of the order of unity.
\bibitem{FOOTNOTE-1} Note that Eq.~(\ref{n_Etyp}) can be written as ${n(E)\sim y}$  in terms of a scaling variable
${y=(E/E_{2/3})^{3/2}}$ introduced in Refs. \cite{silvestrov98,leyronas00} for the IPR
of typical many-body eigenstates, whereas Eq.~(\ref{Ec})  (with the factor $\ln g$ in the denominator) corresponds to the condition $x\sim 1$ in terms of another scaling variable, $x=E/E_1$, used in Refs. \cite{silvestrov98,leyronas00}.
\bibitem{footnote-RMK} In Ref.~\cite{rivas02},
the IPR of a single-particle excitation [hot electron, i.e., our initial state (i)],
as defined by our Eq.~(\ref{IPR-1}), was analyzed. The rescaling of the $y$-axis in
Fig. 9 of Ref.~\cite{rivas02} was chosen to be the same as in Ref.~\cite{leyronas00}
that considered the IPR of many-body eigenstates, our Eq.~(\ref{IPR-2}), which
corresponds to the IPR of a typical basis state [our initial state (ii)], see discussion below Eq. (\ref{IPR-2}).
This choice of the scaling variable $y$, as detailed in Ref. \cite{FOOTNOTE-1}, is indeed appropriate for the setting of Ref.~\cite{leyronas00}, see Eq.~(\ref{Psat-typ}) for $P(\infty)$ in our state (ii).
However, for the analysis of relaxation of a hot electron, which was the subject of Ref.~\cite{rivas02},
the appropriate scaling variable is given by $g^{-1/2}E/\Delta$, as follows from our Eq.~(\ref{Psat-hot}).
The shift of the maximum of the IPR towards $x= 0$ in Fig. 9 of Ref.~\cite{rivas02}
appears to be an artifact of the chosen rescaling of the $y$-axis.
On the other hand, the shift of the point of matching with the golden-rule result towards large $x$, as found in Ref.~\cite{rivas02}, is not affected by the rescaling of the $y$-axis.
\end{thebibliography}
\end{document}